\documentclass[11pt,preprint]{aastex}

\def\la{\raise.5ex\hbox{$<$}\kern-.8em\lower 1mm\hbox{$\sim$}}
\def\ma{\raise.5ex\hbox{$>$}\kern-.8em\lower 1mm\hbox{$\sim$}}

\def\kms{$\rm km\, s^{-1}$}
\def\cm3{$\rm cm^{-3}$}

\def\n0{$\rm n_{0}$}
\def\B0{$\rm B_{0}$}

\def\T r{$\rm T_{ r}$}
\def\T as{$\rm T_{ as}$}

\def\kms{$\rm km\, s^{-1}$}

\slugcomment{Accepted to ApJ}

\shorttitle{The coronal line region in AGN}
\shortauthors{Rodr\'{\i}guez-Ardila et al.}

\begin{document}
\title{Outflows of very ionized gas in the center of  Seyfert galaxies: \\
       kinematics and physical conditions\altaffilmark{1}}

\author{Alberto Rodr\'{\i}guez-Ardila}
\affil{Laborat\'orio Nacional de Astrof\'{\i}sica/MCT, Rua dos Estados Unidos 154, 
CEP 37504-364, Itajub\'a, MG, Brazil}
\email{aardila@lna.br}
\author{M. Almudena Prieto}
\affil{Instituto de Astrofisica de Canarias, Tenerife, Spain \& Max-Planck Institute for Astronomy, Heidelberg, Germany}
\and
\author{Sueli Viegas and Ruth Gruenwald}
\affil{Instituto de Astronomia, Geof\'{\i}sica e Ci\^encias Atmosf\'ericas - USP, Brazil}

\altaffiltext{1}{Based on observations made with ESO Telescopes at the Paranal Observatory under programme ID
68.B-0627}

\begin{abstract}
Medium resolution spectra (100~\kms) are used to deduce the size and
kinematics of the coronal lines in a sample of nearby Seyfert galaxies
(Circinus, NGC 1386, NGC 1068, NGC 3227, NGC 3783 and MCG-6-30-15) by
means of simultaneous observations of [\ion{Fe}{11}]~7889 \AA,
[\ion{Fe}{10}]~6374 \AA, [\ion{Fe}{7}]~6087 \AA, 
[\ion{Si}{6}]~2.48~$\mu$m and [\ion{Si}{7}]~1.96~$\mu$m.  The coronal
lines extend from the unresolved nucleus up to distances    
between a few tens to a few hundreds of parsecs.  The region of the
highest ionized ions studied, [\ion{Fe}{11}] and [\ion{Fe}{10}], is the least
spatially extended, and concentrates at the center; intermediate
ionization lines extend from the nucleus up to a few tens to a few hundred
parsecs; lower [\ion{O}{3}]-like ions are known to extend in the kpc
range. All together indicates a stratification in the ionized gas,
usually interpreted in terms of nuclear photoionization as the driving
ionization mechanism.  However, coronal
line profiles show various peculiarities: they are broader by a
factor of two than lower
ionization lines, the broadening being in terms of asymmetric blue
wings, and their centroid position at the nucleus is blueshifted by a few
hundreds of \kms. Moreover, in two objects --NGC~1386 and NGC~1068,
a double peak [\ion{Fe}{7}]~6087~\AA\ line is detected in the nuclear
and extended coronal region, this being the first report in the
literature of such type of profile in coronal lines in active galactic
nuclei. If interpreted as outflow signatures, the total broadening 
of the lines at zero intensity levels implies gas velocities up to 2000~\kms.
Although the stratification of ions across the coronal region means
that photoionization is the main power mechanism, the high
velocities deduced from the line profiles, the relatively large
spatial extension of the emission, and the
results from single cloud photoionization models indicate  that an
additional mechanism is at work. We suggest that shocks generated by the
outflow could provide the additional required power for line
formation.  

\end{abstract} 


\section{Introduction}

Coronal lines (CL) are collisionally excited forbidden transitions
within low-lying levels of highly ionized species (ionization 
potential $>$ 100
eV). They can be produced by either a hard UV continuum \citep{fer97}, 
a hot collisionally ionized plasma \citep{vac89}
or most plausible, a mix of both processes \citep{con98}. 
Because of their high ionization potential (IP), these lines
are particularly suitable for getting information on the otherwise 
difficult to access observationally, extreme UV - soft X-ray region of
the ionizing spectrum \citep{pr00}.  Due to the high energies
involved in their production, coronal lines are genuine tracers of the
AGN power mechanism.  Pure starbursts when boosted by massive O-B 
stars in their Wolf-Rayet phase, may  show \ion{He}{2} lines but
not lines of higher ionization potential ions ($\chi >54$~eV).

Observationally, coronal lines are present with comparable strength in
AGN spectra regardless of their class \citep{pe84,er97,pr00,rod02,reu03}. 
The peak position is usually blueshifted relative to the systemic
velocity of the host galaxy ($\Delta$V $\sim$ 500-800~\kms) and
broader than low ionization lines \citep{pe84,dro86,er97,rod02}.  
That has led to the idea that CL are associated with outflows,
formed in a intermediate region between the classical narrow-line- and
broad-line-region (NLR and BLR, respectively).  The broad line region
has sizes of less than a few hundred light-days \citep{pet04}
and thus remains unresolved; the coronal region because of its lower
velocity dispersion should be placed further out but still relatively
close to the BLR, considering the high ionization level of the gas.
Accordingly, the coronal region has also remained spatially
unresolved. 

The physical nature of the CLR has called the attention of many
authors starting by the pioneer work by \citet{pe84}, who
discover the systematic blueshifting and broadening of this high
ionization gas.  \citet{er97} found that the CL occurs
predominantly in objects with a soft X-ray excess and suggest a
possible relationship between the CLR and the absorption edges
associated with warm absorbers present in nearly 50\% of AGN.  \citet{por99}
showed that the CL may, in fact, be formed by the warm
absorber.  If that is the case, the CLR should be dense
($n\sim$10$^{6}$ cm$^{-3}$) and located at comparable distance as the
broad line region.  On the other hand, \citet{fer97}
photoionization calculations show that the CLR can form in gas with
densities of $ 10^2 - 10^{8.5} $ cm$^{-3}$ and extends up to several
hundred parsecs, thus coinciding with the size of the extended
NLR. However, they claim that except for the lower ionization
coronal lines ([Ne\,{\sc v}], [Si\,{\sc vii}], [Ca\,{\sc viii}] and
[Fe\,{\sc vii}]) that can form efficiently in gas that is roughly
$\sim$10~pc and beyond, the rest of coronal lines form optimally in
gas that is less than 10~pc from the ionization source.

Coronal lines are present all across the electromagnetic spectrum. The
highest ionization species are located in the X-rays region. In the optical,
the most prominent ones are from iron. Dust extinction 
and the fact that reliable detection of these lines requires medium to
high spectral resolution and large S/N ratios has made  
measurements of these lines difficult.  The near- and mid-infrared
regions, less affected by extinction, are very rich in coronal lines
from different species and ionization levels.  Indeed, they are often
the dominant lines in the near-IR spectra of Seyfert galaxies
\citep{reu03}.

The current availability of adaptive optics in 8 m
telescopes enabled \citet{pri05} to determine the size and
morphology of the coronal line region for the first time. That is
found to extend from the nucleus up to 30 to 200~pc radii at most, and
be aligned preferentially with the direction of the lower-ionization
cones seen in Seyfert galaxies. As expected, the coronal line region (CLR) is primarily 
produced very close to the
active nucleus but is nevertheless resolved over several tens of
parsecs.  This, together with its preferential alignment with the
ionization cone prompts to the possibility that the nuclear radiation
is intrinsically collimated or that some additional in-situ excitation
mechanism is needed to explain CL gas at those observed distances from the
ionizing source. This paper attempts to address those issues by
studying the physical conditions of the coronal line gas and its
kinematics.

To that aim, we present  high-sensitive observations with spectral
resolution R$\sim$100 km~s$^{-1}$ of Fe and Si coronal lines located
in the visible and near-infrared spectra respectively. As stated
above, the goal is determining the kinematics of the emitting region
and constraining the physical properties of the coronal gas. To that
aim, a sample of six nearby AGN were selected, and spectra centered at
the position of a selected group of bright coronal lines are studied.
In the optical, we selected [Fe\,{\sc vii}]~6087~\AA\ (IP=100~eV), 
[Fe\,{\sc x}]~6374~\AA\ (IP=240~eV) and
[Fe\,{\sc xi}]~7889~\AA\ (IP=260e~V). In the near-IR, [Si\,{\sc
vi}]~1.963~$\mu$m (IP=170~eV) and [Si\,{\sc vii}]~2.48~$\mu$m
(IP=205~eV) were chosen. These lines are all located near in
wavelength to low-ionization lines, easing the comparison regarding 
the size of the emitting region and the shape of  the line profiles
between different ionization levels. In \S~\ref{obs}, we
describe the sample, the observations and data reduction process.  The
determination of the coronal region size based on the available data is
carried out in \S~\ref{size}. In \S~\ref{comp} we discuss and compare
the size of the CLR with that determined for the NLR using our data
and the literature.  A comparison of the observed emission line
profiles to unravel the kinematics of the coronal gas is in
\S~\ref{kin}. Photoionization and shock models to explain the physical
conditions of the coronal gas are described in \S~\ref{mod}. An
overview of our main results and conclusions are in \S~\ref{con}.

\section{Observations} \label{obs}

In order to maximize the spatial resolution in physical scales, the sample of
objects chosen for this work is composed of six of the nearest and
brightest Seyfert galaxies: Circinus (Seyfert~2), NGC\,1068 (Seyfert~2), 
NGC\,1386 (Seyfert~2), NGC\,3783 (Seyfert~1), MCG-6-30-15 (Seyfert~1) 
and NGC\,3227 (Seyfert~1.5). Table~\ref{pa} lists the
characteristics of these galaxies. The spatial scale (pc/$\arcsec$)
was derived assuming H$_{\rm o}$=75~km~s$^{-1}$~Mpc$^{-1}$. These
objects were observed at medium resolution spectroscopy, R=2600
($\sim$120 km~s$^{-1}$) in the optical, R=3000 ($\sim$100 km~s$^{-1}$)
in the near-IR, with a slit width=1$\arcsec$, the slit being always
oriented north-south.  The slit was nearly aligned to the PA of the
[\ion{O}{3}] emission for most sources, listed in column~5 of
Table~\ref{pa}.  The differences in position angle of the extended
ionized gas and that of the slit is, for the majority of sources, not
larger than 15\arcdeg, except for MCG-6-30-15, where the difference is
65\arcdeg and the opening angle is small, about 30\arcdeg maximum.  In
Circinus, the slit is aligned with the edge of the ionization cone:
the PA of the cone is -44\arcdeg NW but the opening angle of the cone
is $\sim$90\arcdeg.

The spectroscopic configurations, telescopes and instrumentation 
employed for the optical and near-infrared observations as well as 
data reduction and extraction procedures were as follows.  Log of the 
observations is in Table~\ref{log}.

The optical iron coronal lines [Fe\,{\sc vii}], [Fe\,{\sc x}] and
[Fe\,{\sc xi}] were observed, in service mode (program 68.B-0627(A)), 
with the imaging-spectrograph EMMI at the ESO/NTT. 
Two grating positions were needed: one centered at
$\lambda$=6230 \AA, in order to include the [Fe\,{\sc vii}] and
[Fe\,{\sc x}] lines and the low-ionization line [O\,{\sc i}]
$\lambda$~6300~\AA, used for comparative purposes. The plate scale of
this setup is 0.35$\arcsec$/pix. The second position was centered at
$\lambda$=7950 \AA\ to observe the [Fe\,{\sc xi}] $\lambda$7890~\AA\
line (plate scale 0.27$\arcsec$/pix).

Data were reduced using {\it IRAF} standard procedures, that is, bias
subtraction and division by a flat field frame. Wavelength and flux
calibration were carried out by means of HeNeAr lamp frames and
observations of spectroscopic standards taken during the night,
respectively.  A log of observations is presented in Table~\ref{log}.
1-D frames were extracted from the 2-D images by summing up rows along
the spatial direction. The optimal number of rows to be summed varied
from source to source and depended on the achieved spatial resolution.

For the near-IR coronal lines, the observations were carried out 
also in service mode (program 68.B-0627(B)) with
the imager-spectrograph ISAAC mounted on the VLT/ANTU 8.2m telescope.
Two different grating positions were used.  For the spectral region
covering [Si\,{\sc vi}]~19630~\AA, the grating was centered at
$\lambda$=19800\AA. This setup allows the simultaneous observation of
that line and H$_{2}$~19570~\AA.  The second grating position,
centered at 24400~\AA, was to observe [Si\,{\sc vii}]~24800~\AA\ and
the Q(0) H$_{2}$ lines. The detector was a 1024$\times$1024 array with
a spatial scale of 0.148$\arcsec$/pix and a spectral resolution of
6.5~\AA\ FWHM.  In order to assure good sky subtraction, the targets
were observed following the classical ABBA technique. Data were
reduced using {\it Eclipse} and {\it IRAF} routines. First, we
corrected for the ``electrical ghost'' features generated by the
detector using the {\it Eclipse} package. The data were then
flat-fielded using a master flat field image following a correction
for distortions along and perpendicular to the slit. Wavelength
calibration was performed by means of bright OH lines present in the
frames. The 1-D spectra were extracted following a similar procedure
as that described for the optical data.  1-D spectra were then
corrected for telluric absorptions using a telluric standard star
observed right before/after the science target. Airmass differences
between the standard star and the objects was usually 0.1 or less,
except for the case of Circinus, where the airmass difference of the
observation centered at 24400~\AA\ was $\sim$0.25.  Flux calibration
was applied to the data assuming that the observed telluric star
continuum was well approximated by a blackbody function of temperature
and magnitude equal to that of the telluric star. The derived
sensitive function was then applied to the galaxies.

The intrinsic spatial resolution of the optical and NIR spectra
was dictated by the seeing at the time of the different integrations.
It was determined from the FWHM of the light profile of the standard star
observed right before or after the target.  This estimate is listed in
column~5 of Table~\ref{log}. The spectral resolution is 
$\sim$100~\kms\ in the optical, $\sim$100 \kms\ at 1.96$\mu$m, and 
$\sim$78 \kms\ at 2.48~$\mu$m.  The RMS of the wavelength calibration is 
3~\kms\ for the optical spectra centred at 6230~\AA, 1.3~\kms\ for the 
ones centred at 7950~\AA\ and 4.6~\kms\ for the NIR data. 
The spatial resolution in the NIR is FWHM $\simeq$ 0.75" and
in the optical $\sim 1 \arcsec$.

\section{Results} \label{size}

\subsection{Spatial extraction of the coronal line spectra}

2-D spectra of each galaxy were averaged in the spatial direction into
separate extraction windows. The spatial size of these windows were
selected to optimize both signal-to-noise and spatial resolution. The
sizes of the extraction window are given in Table~\ref{log}, column~7, and in
general correspond to the achieved spatial resolution (seeing) of the data.  In
galaxies with strong continuum, to mitigate nuclear scatter light in
the off-nuclear spectra, the nuclear extraction window was enlarged by
factors between 20\% and 100\%.

Aperture windows were extracted consecutively north and south the
nuclear one.  Because of the differences in plate scale between the
blue and red detectors in the optical and the near-IR array, besides
the effect of seeing variations, the extraction window is different
from line to line even within the same object.

Figures~\ref{fig1} to~\ref{fig6} show the extracted spectra along the
spatial direction for all the objects in the sample. Usually, one or
two extractions ahead of the last one showing extended coronal
emission are plotted. Each spectrum is identified by its distance (in
parsecs) to the nucleus, measured from the center of the extraction
window to the nuclear peak position. The value following the $\pm$
sign is the radius of the extraction window.  The dotted lines mark
the position, relative to the systemic velocity of the galaxy, of
[Fe\,{\sc vii}]~6087~\AA\ and [Fe\,{\sc x}]~6374~\AA\ (blue spectra)
and [Fe\,{\sc xi}]~7889~\AA\ (red spectra).  [Fe\,{\sc vii}]~5721~\AA\
is also present in all objects. This line is weaker than [Fe\,{\sc
vii}]~6087~\AA\ and therefore is not shown in any of the figures.

Information about near-infrared coronal lines [Si\,{\sc
vi}]~19630~\AA\ and [Si\,{\sc vii}]~24830~\AA\ is available for three
objects in the sample: Circinus, NGC~1068 and NGC~3727. No ESO/ISAAC
observations of the remaining galaxies could be scheduled within
ESO/Paranal service mode observations.  Spectra for those three
sources are shown in Figures~\ref{fig7}, ~\ref{fig8} and~\ref{fig9},
respectively. The spatial resolution $\simeq 0.6''$ FWHM is superior
than that achieved in the optical; the spectral resolution is
comparable, 100~\kms.

Table~\ref{sizeclr} lists the size (radius, in parsecs) of the coronal region
derived from [Fe\,{\sc vii}]~6087~\AA, [Fe\,{\sc x}]~6374~\AA\ and
[Fe\,{\sc xi}]~7889~\AA, respectively.  For comparison, the table also
lists the radius of the emitting region measured from [O\,{\sc i}]~6300~\AA\ 
and [\ion{S}{3}]~6312~\AA.  The size derived from the
latter is uncertain because the line is intrinsically weak outside the
nuclear region. The last column in the table gives the NLR radius
based on [O\,{\sc iii}]~5007~\AA\ line emission data taken from the
literature.

Tables~\ref{fwhmopt} and~\ref{fwhmnir} list the FWHM (in \kms,
already corrected for instrumental resolution) of the coronal lines as well as
that of [\ion{O}{1}]~6300~\AA, included for comparative purposes. Also, 
the shift of the centroid position of each line component relative to
the systemic velocity is given. Centroids and FWHMs were
estimated after a multi-Gaussian fit to the lines. In most cases,
a two Gaussian fit, accounting for a narrow and a broad component, 
was necessary to improve the reduced $\chi^2$. In cases of low 
signal-to-noise, only a single component fit was used.
The tables provide the resulting parameters from a two Gaussian fit when that 
was possible. In [\ion{Fe}{10}]~6374~\AA\, the fitting procedure accounts  
for the presence of [\ion{O}{1}]~6363~\AA, whose width and flux 
were fixed according with the values derived for [\ion{O}{1}]~6300~\AA.
Typical errors in the Gaussian centroid position was $\sim$20~\kms.

\subsection{Specifics to each object}

\subsubsection{Circinus}

\begin{figure}
\epsscale{1.1}
\plotone{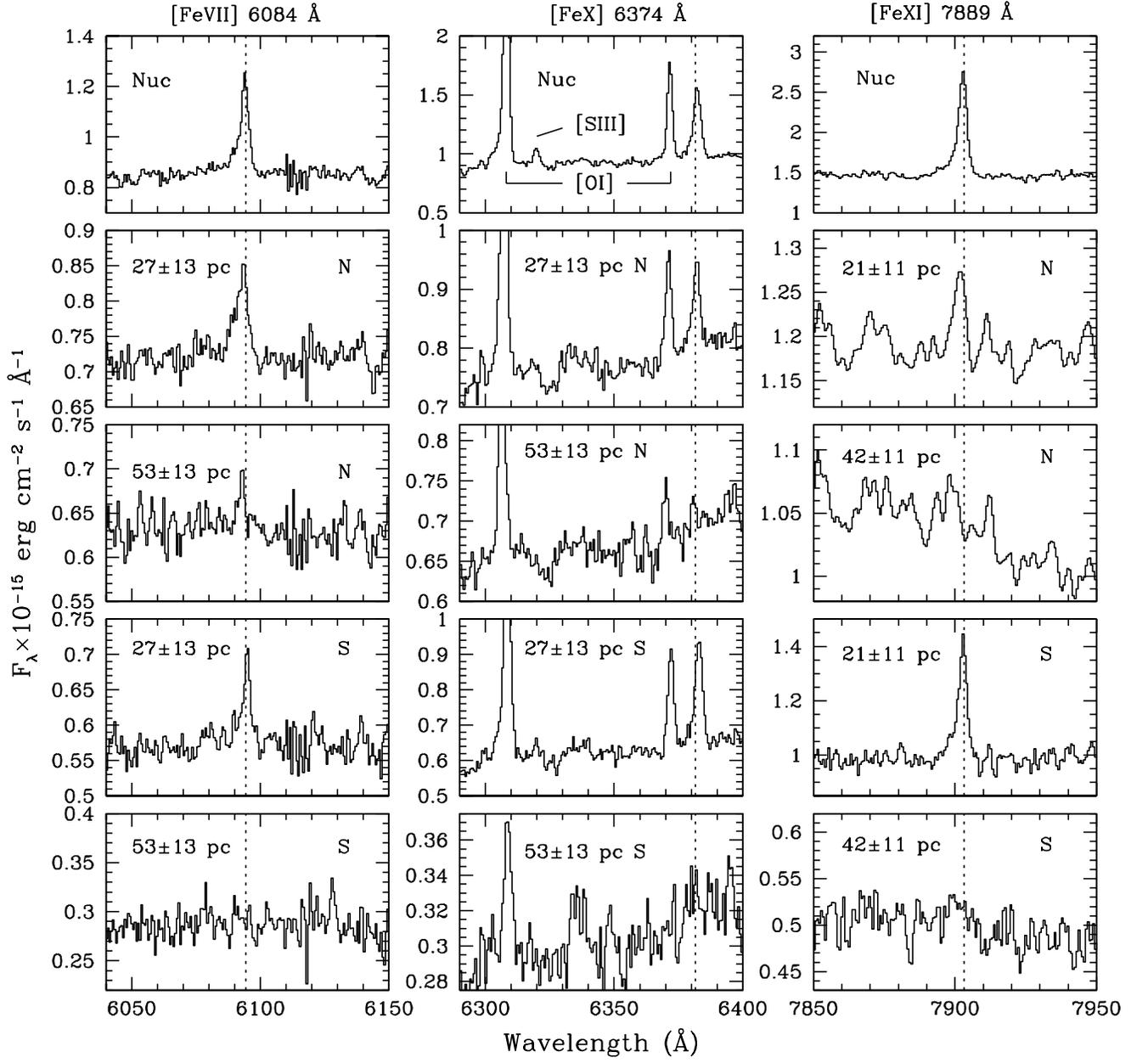}
\caption{NTT  spectra of Circinus along the spatial direction showing the extent of the [Fe\,{\sc vii}]~6087~\AA\ (left panels),
[Fe\,{\sc x}]~6374~\AA\ (middle panels) and [Fe\,{\sc xi}]~7889~\AA\ (right panels). The spectra were extracted according to
the aperture size D   indicated in Table 2, and spatially
correspond to a mean distance to the nucleus indicated in each box. The orientation of the slit is always
north-south.
The dot-line marks the systemic velocity at each of the three lines. \label{fig1}}
\end{figure}

\begin{figure}
\epsscale{1.1}
\plotone{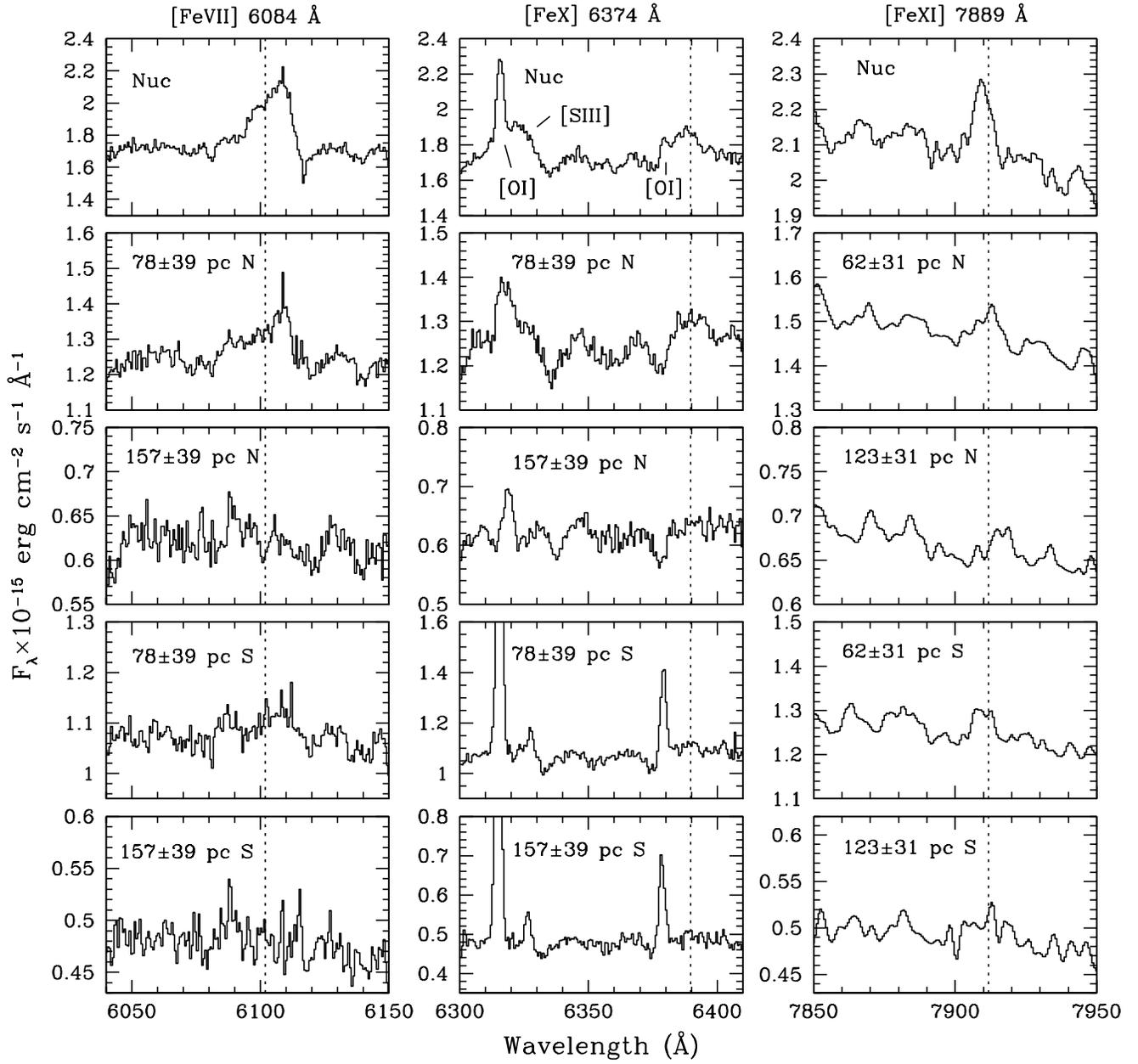}
\caption{The same as Fig~\ref{fig1} for NGC~1386.  \label{fig2}}
\end{figure}

This is the nearest AGN in the sample, allowing us to map the gas
emission with a spatial resolution  of $\sim$27~pc
in the blue and $\sim$20~pc in the red.  In the optical spectra
(Fig.~\ref{fig1}), [Fe\,{\sc vii}]~6087~\AA, [Fe\,{\sc x}]~6374~\AA\ and
[Fe\,{\sc xi}]~7889~\AA\ are prominent and broader than
[\ion{O}{1}]~6300~\AA. All the iron line regions extend several tens of
parsec, including [\ion{Fe}{11}], which extends within the central 20~pc
radius (Table~\ref{sizeclr}).  In contrast, [\ion{O}{1}]~6300~\AA\ extends up to
200~pc radius (Figure~\ref{fig1} shows only the central 53~pc region).

The Fe coronal lines are stronger in the north direction, coinciding
with the location of the ionization cone. South of the nucleus, the
line spectrum is much weaker due to a combination of extinction by the
large dust lanes that cover the southeast region and the presence of
a circumnuclear star forming region.  This is supported by the
weakness of [\ion{O}{1}]~6300~\AA, which is almost reduced to
noise at 100~pc south from the center and the line ratios
[\ion{N}{2}]/H$\alpha$ and [\ion{S}{2}]/H$\alpha$, almost a factor of
two smaller than those at the same distance to the north.

The Si coronal lines [Si\,{\sc vi}] and [Si\,{\sc vii}] extend to
similar distances but the extension is seen
equally north and south the nucleus; [\ion{Si}{7}] extends further
north, up to 35~pc (Fig.~\ref{fig7}).  The inferred size of [\ion{Si}{7}] region
confirms VLT subarcsec imaging results in this line \citep{pri05}.  
Indeed, the [\ion{Si}{7}] image reveals a two-side ionization
cone, the North-West  side being located  about 30~pc length within the much
larger, one-side only, cone seen in
[\ion{O}{3}]~5007~\AA\ or H$\alpha$.

 \citet{ol94,ol99} report for Circinus coronal lines widths
 comparable to those of lower ionization lines, FWHM $\sim$100~\kms.
 As illustrated by Figs.~\ref{fig1} and~\ref{fig7}, all the coronal lines show an
 asymmetric profile with a sharp cut-off at the red side and an
 extended wing to the blue side - less obvious, e.g., in
 [\ion{O}{1}]~6300~\AA.  The FWHM of both low and high ionization
 lines are rather similar, $\sim$100~\kms; yet, at one-third of the
 line maximum, all coronal lines  have developed a
 prominent blue-asymmetric wing, and at 10\% of the maximum, this wing
extends to $\sim$400~\kms\ at the nucleus.  At this same position to the red, the
 highest velocity measured is 200~\kms\ 
(the nuclear profile in velocity space is shown in Fig.~\ref{fig11}).  The same
 characteristics apply to the Si line profiles.
The shape and width of the lines are in excellent agreement with those
presented by \citet{mus06} on the basis of much
higher spatial resolution data obtained from the near-IR integral
field spectrograph SINFONI at the VLT. Their nuclear spectrum, which covers a
region of $\sim$15~pc displays prominent [\ion{Si}{6}]~1.963~$\mu$m,
[\ion{Al}{9}]~2.040~$\mu$m and [\ion{Ca}{8}]~2.32~$\mu$m lines, all
showing the same kinematics characteristics, namely, a blue-shifted
wing reaching 500~\kms\ at the base of the line but only 200~\kms\ on 
the red side (cf. their Figure~8).
                                                                         
As a reference, H$_{2}$ molecular lines - seen in large number in the
ISAAC spectra - are detected up to 150~pc from the centre (only the
central 51~pc region is shown in the figures). H$_{2}$ emission is the
strongest at the nucleus, and then fades progressively outwards. At
about 20~pc south from the nucleus H$_2$ is slighted enhanced due
to the presence of a star forming ring.

\subsubsection{NGC\,1386}

This object is a factor 2 farther than Circinus. The coronal
emission  was  mapped with a spatial scale of $\simeq$70~pc FWHM (Table~\ref{log}).

Fig.~\ref{fig2} shows weak extended coronal emission up to 50~pc radius in 
[\ion{Fe}{10}] and [\ion{Fe}{11}], and up $\sim$ 100~pc radius in [\ion{Fe}{7}] (
 see also Table~\ref{sizeclr}).  The extended emission is slightly stronger north
 of the nucleus than to the south. In contrast, [\ion{O}{1}] and [\ion{S}{3}] tend
 to be stronger towards the south. [\ion{O}{3}]~5007~\AA\ shows high
 collimated emission, composed of several blobs distributed in the
 north-south direction \citet{sch03}.  The brightest [\ion{O}{3}]
 blob is about 1~$\arcsec$ north, coinciding with extended [\ion{Fe}{7}] at
 the same location in the spectra.

The line profiles in the nuclear region are complex: low and high
ionization lines show a broad  component but  [\ion{O}{1}] shows in 
addition a prominent narrow
component, FWHM $\simeq$100~\kms, whereas [\ion{Ar}{3}] and the Fe coronal lines show
a broader profile: 300~\kms $<$ FWHM $<\sim$ 800~\kms.  [\ion{Fe}{7}]~6087~\AA, 
having the highest S/N, shows a double peak profile, with the red peak
stronger than the blue one.  This double peak
is seen up to $\sim$80~pc north and south of the nucleus. 
A double peak profile is also seen in [\ion{Fe}{7}]~5721~\AA, which is 
detected in the spectra but is not shown because it is
weaker than [\ion{Fe}{7}]~6087~\AA.  As reported
below, NGC~1068  presents a similar double peak profile in [\ion{Fe}{7}].
The signal-to-noise limits the discussion about the 
[\ion{Fe}{10}] and [\ion{Fe}{11}]
profiles, where only a single Gaussian could be fitted to the data. 
No NIR coronal line observations are available for this object.

\subsubsection{NGC~1068}

\begin{figure}
\epsscale{1.1}
\plotone{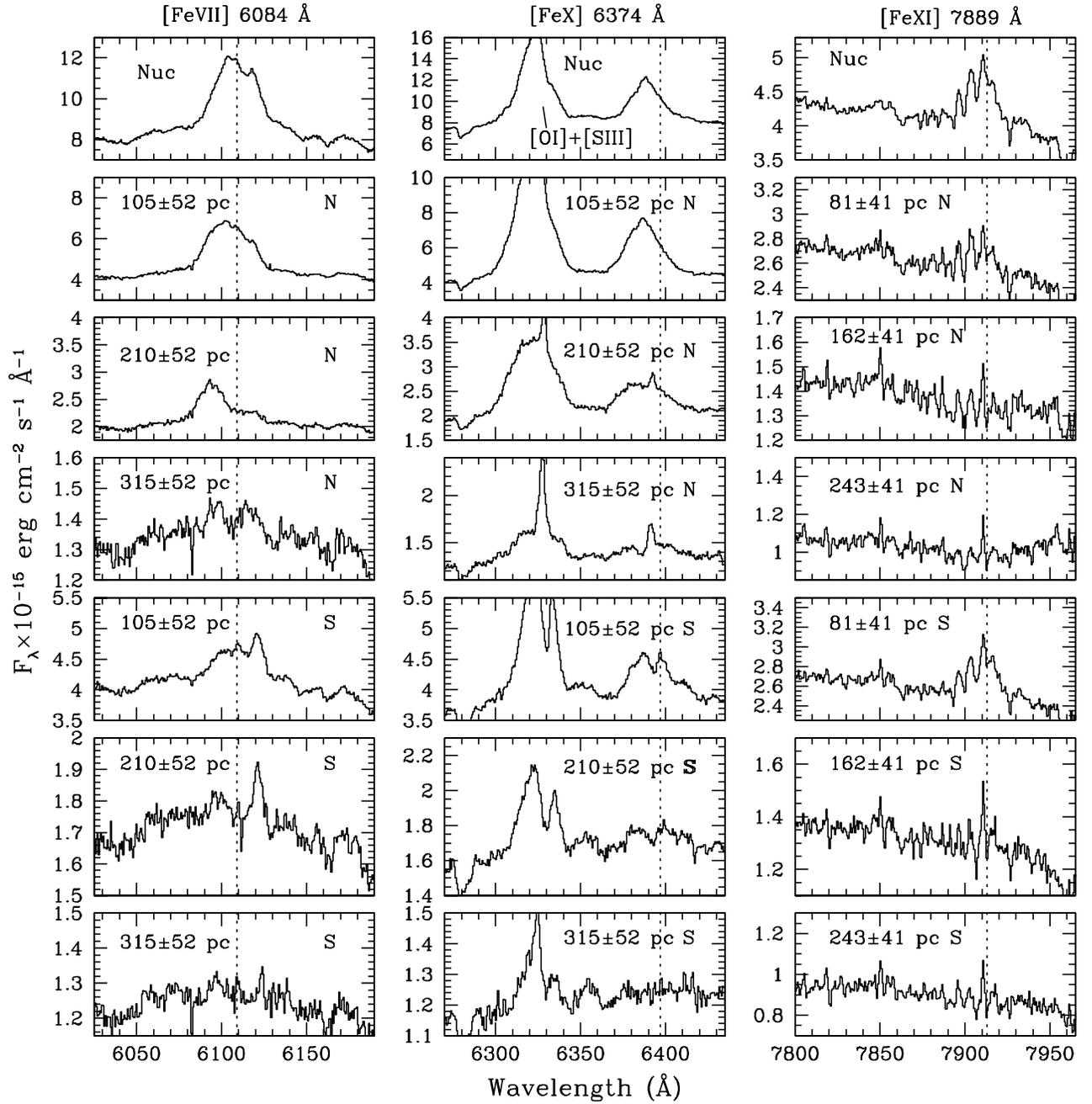}
\caption{The same as Figure~\ref{fig1} for NGC~1068. In all positions, the presence
of [\ion{Fe}{10}]~6374~\AA\ was determined after a consisting multigaussian fit to the region 
[\ion{O}{1}]+[\ion{S}{3}] -- [\ion{O}{1}]+[\ion{Fe}{10}].
The strong line seen  starting from 105~pc to the south, redwards of [\ion{O}{1}]~6300, 
is presumably [\ion{S}{3}]~6312~\AA.
Wrinkles on [\ion{Fe}{11}] line are telluric residuals. \label{fig3}}
\end{figure}

NGC~1068 is at nearly the same distance as NGC~1386 and NGC~3727. The
spatial scale achieved in this case is $\sim$100~pc in the blue
spectra and $\sim$80~pc in the red one.  A rapid inspection to the optical
spectra (Fig.~\ref{fig3}) reveals its complexity, dominated by
extreme broadening in all, low and high ionization lines.  Even at
our mid-spectral resolution, it is difficult to characterize 
the line profiles and their spatial extension due to blending.  For example, the 
FWHM of [\ion{Fe}{7}] is $\simeq$ 1700~\kms\ whereas that of 
[\ion{O}{1}] is $\sim$1000~\kms. NGC~1068 shows
thus the largest gas velocities in the sample.  Hence, most of the
coronal emission discussion is limited to [\ion{Fe}{7}]~6087~\AA\ and 
[\ion{Si}{7}]~2.48~$\mu$m since these are the best isolated lines.

[Fe\,{\sc vii}]~6087~\AA\ is the most extended CL, up to 210~pc
north to the nucleus, but [\ion{Si}{6}] -- next in ionization potential --
is by far the strongest
line, a factor 3 stronger, although covering half that spatial size.  All other
coronal lines decrease in strength and size of the emitting
region with
increasing IP, [\ion{Fe}{11}] being the weakest, a factor 30 lower in flux
than [\ion{Si}{6}], and the smallest region, less than 40~pc radius (Tables~\ref{fwhmopt}
and~\ref{fwhmnir}).  The characterization of the [\ion{Fe}{10}]~6374~\AA\ region is
difficult because this line is strongly blended with
[\ion{O}{1}]~6363~\AA. A multiple Gaussian fitting, fixing parameters
for [\ion{O}{1}]~6363~\AA\ as determined from a fit to the 
[\ion{O}{1}]~6300~\AA\ line, leads to a conservative upper 
limit of 100~pc radius for the [\ion{Fe}{10}]~6374~\AA\ region.

Both [Si\,{\sc vi}]~19630~\AA\ and [Si\,{\sc vii}]~24830~\AA\ extend
north and south of the nucleus up to 100~pc radius
(Fig.~\ref{fig8}), half the [\ion{Fe}{7}] region but comparable to that
of [Fe\,{\sc x}] (Table~\ref{sizeclr}).  The particular spatial morphology of [Si\,{\sc vii}]
emission region has been revealed at subarcsec scales by VLT adaptive
optics images \citep{pri05}. [Si\,{\sc vi}] shows a central three blob
structure within the inner 7~pc radius surrounded by diffuse
emission up to $\sim$70~pc radius.  The present spectroscopy just
confirms the Si coronal size derived from imaging.  The reported size
for the [Si\,{\sc vi}]~19630~\AA\ region, derived from {\it NICMOS}
imaging \citep{tho01} is much larger: $\sim$300\,pc to the
south and 200\,pc to the north. However, we believe that Thompson's et
al. image is largely contaminated by the satellite line
H$_{2}$~19570~\AA.  This is easily seen in Figure~\ref{fig8}, where
both lines appear heavily blended.  Moreover, Figure~\ref{fig8} shows
that at distances larger than $\sim$135\,pc, only H$_{2}$ is detected.

For comparison, [O\,{\sc i}]~6300~\AA\ is detected up to 525~pc north
and south from the centre (Figure~\ref{fig3} shows a smaller region).
All lines are stronger towards the north side of the nucleus, which
coincides with the preferential direction of the [\ion{O}{3}] line
emission and the radio jet.

Undoubtedly the most reliable information about the kinematics of the coronal gas
can be derived first from [\ion{Fe}{7}]~6087~\AA, which is the best isolated line and
second, from [\ion{Si}{7}]~2.48$\mu$m, also isolated but just falling at the
edge of the ISAAC detector, hampering the detection of the most
redshifted gas.
  
At the unresolved nucleus, [\ion{Fe}{7}]~6087~\AA\ shows a prominent
double-peak profile (FWHM $\sim$ 1700 \kms)  on top of a much
broader component with FWZI $>$4000 \kms.  The spatially
resolved [\ion{Fe}{7}] also shows a double peak component at both north and
south of the nucleus (Fig.~\ref{fig3}). At first glance, it could be
said that the double peak may result from the combination
of a large integration window size and a strong velocity
gradient. In order to confirm that this peculiar profile structure is real,
we have extracted spectra row by row along the spatial direction
in the 2-D frame of NGC~1068. At the redshift of this galaxy (see Table~\ref{pa}), 
each pixel corresponds to a projected distance of about 26~pc. 
The left panel of Figure~\ref{fig10} shows the resulting spectra mapped up to 
250~pc north and 180~pc south of the peak light profile distribution.
The fact that the double peak profile in [\ion{Fe}{7}] is 
detected at scales below the one covered by the integrated
spectra of Fig.~\ref{fig3} strongly supports the
presence of double peaked coronal lines in NGC~1068. 
Moreover, it can be seen that the relative distance 
between the blue and red peaks 
increases outwards to the north, from $\sim$700~\kms\ at the
nucleus to $\sim$1400~\kms\ at 180~pc. Note, however, that the shift of the 
red peak changes little relative
to the line centroid. It moves from $\sim$500~\kms\ in the nucleus
to $\sim$600~\kms\ at 180~pc. In contrast, the blue peak shifts from
$\simeq$ -200~\kms\ to ~$\simeq$ -800~\kms\ within the same distance interval.
Further to the north, the centroid position of each of the two peaks
remains stable, with the red peak being significantly weaker than the blue
one. South to the nucleus, the shape of the coronal profile 
changes significantly if compared to the northern one: the
blue peak becomes broader and its centroid position gets closer to 
the red one. At $\sim$100~pc south from the centre, the blue and red peak have a FWHM of 
$\sim$1600~\kms\ and $\sim$300~\kms, respectively, but the former 
starts fading gradually. In fact, at 180~pc, only a narrow red peak
is visible. 

The right panel of Fig.~\ref{fig10} shows equivalent
line profiles for  [\ion{O}{1}]~6300~\AA, emphasizing the
different kinematics followed by neutral gas.
Unfortunately, [\ion{O}{1}]~6300~\AA\
is polluted by [\ion{S}{3}]~6312~\AA, which is separated from the former, in
velocity space, by $\simeq$570~\kms\ at rest, which
makes uncertain its interpretation. 
Nevertheless, two main differences arise: first, no evidence of a double peak
structure is seen at the nucleus. Second, the centroid 
position of the line coincides with the systemic velocity of the galaxy 
and its shift along the spatial direction is consistent 
with rotation around the nucleus. The only similarity seen between
both profiles is the presence of a narrow read peak that appears at $\sim$
50~pc to the south and $\sim$130~pc to the north. The peak to the north is 
probably a second narrow [\ion{O}{1}] component, judging from a similar 
component seen in [\ion{O}{1}]~6363~\AA\ (see Fig.~\ref{fig3}), 
the component to the south is presumably due to [\ion{S}{3}].

For comparison, HST/STIS spectroscopy shows
complex [\ion{O}{3}]~5007~\AA\ and H$\beta$ line profiles \citep{cec02},
with multiple broad peak components and persistent line blueshifting,
indicating velocities larger than 2000~\kms\ at the faintest levels.
These complex profiles have their origin in a multitude of knots,
many unresolved (size $<$ 4~pc), and filamentary structure
seen across the extended narrow line region of the galaxy (cf. HST/FOC
[\ion{O}{3}]~5007~\AA\ images of NGC~1068). The spatial resolution of 
the present spectra is poorer, very likely averaging over 
many of the clouds seen in the FOC
image, including mostly those along the jet.  However, in velocity space, the
[\ion{Fe}{7}] line traces the same complex profile and gas velocities
(Fig. 10). Thus, the [\ion{O}{3}] and [\ion{Fe}{7}] clouds should be the 
same.

Summarizing, the high ionization gas, as mapped by [\ion{Fe}{7}], 
shows clear imprints of an outflowing wind, with  the approaching and the 
receding components visible at different spatial locations, and their speed 
increasing outward. In contrast, the neutral gas, as mapped by [\ion{O}{1}],
seems to have a more relax kinematics.

Moving up in IP, the line profiles become narrower (Figs.~\ref{fig3} 
and~\ref{fig8}). Gas velocities traced by [\ion{Fe}{11}] (although the profile is badly affected
by telluric absorptions a broad Gaussian fit is feasible) and the silicon
lines are in the range 300~\kms\ $ < $ FWHM $<$ 1000~\kms.  The 
non-detection of higher velocity components is a limitation of the spatial
resolution and signal-to noise: higher velocities components are
dumped in the continuum noise.  The Si lines follow a similar
kinematic as that shown by [\ion{Fe}{7}]: the peak position of the silicon
lines are blueshifted to the north and redshifted to the
south. However, the profiles do not show the prominent double peak
seen in [\ion{Fe}{7}], which is probably because the ISAAC spectra have
higher spatial resolution, hence the possibility to spatially separate
the blue- and red-shifted components. Still, at  certain locations
they do show hints for multiple velocity components (e.g. [\ion{Si}{6}] at
45~pc south; [\ion{Si}{7}] at 135~pc north, Fig.~\ref{fig8}).  Overall, the Si and
[\ion{Fe}{7}] line profiles
show a broader profile  north of the nucleus than  
south of it (Figure~\ref{fig8}); since the radio jet is seen  north, 
the broadening might be associated  with the pass of the jet through the gas clouds. 

Previously, \citet{marc96} reported FWHM $\sim$1000~\kms\
for both optical and NIR
coronal lines, which is lower than those measured in
intermediate ionization lines, e.g. [\ion{O}{3}]~5007~\AA\ (FWHM
$\sim$1500~ km/s), in apparent contradiction with the present results. 
They also reported on systematic line blueshifts that
increase with the IP, reaching $\sim$ 300~\kms, 
but such trend is not obvious from the present data.

\subsubsection{NGC~3227}

This is a Seyfert 1.5 at the same distance as NGC~1068.  The spatial
 resolution achieved was $\sim$60~pc in the blue, $\sim$40~pc in the
 red. However, to avoid contamination of scattered light from
 the broad-line region -traced by H$\alpha$ -- in the off-nuclear
 spectra, a larger extraction window, with diameter of 100~pc in the blue,
 80~pc in the red was considered.

\begin{figure}
\epsscale{1.1}
\plotone{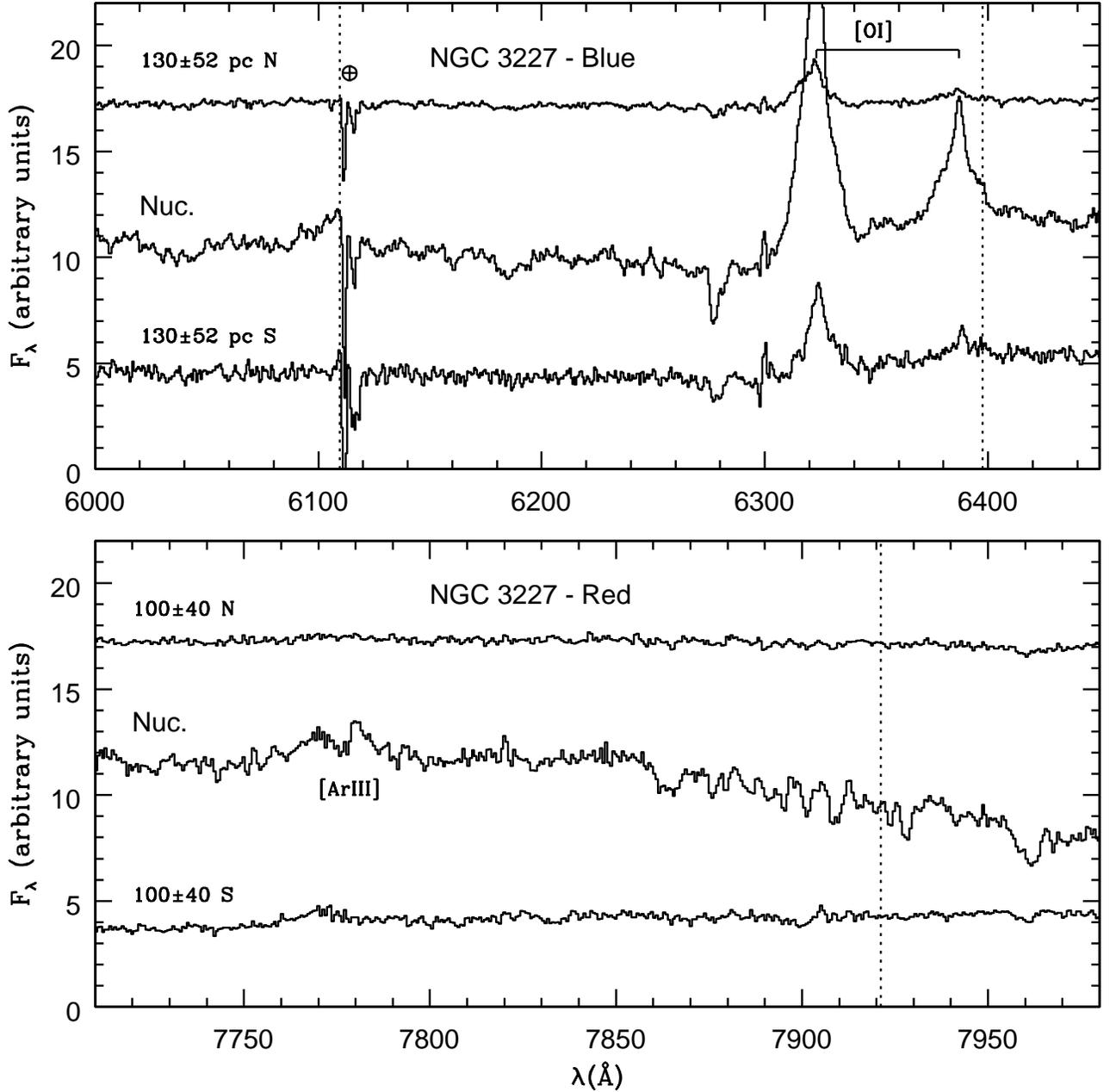}
\caption{Extracted spectra for NGC~3227. The relative
  large size of the nuclear
  aperture used in this case is dictated by  the number of spectra affected by
  scattered light by the  broad line region  - traced by H$\alpha$ in our
  spectra. This is to warrant the real spatial extension of 
coronal emission.  The [\ion{Fe}{7}] profile is
  affected by sky residuals. This Seyfert type 1.5 is the only case  in the sample 
lacking photons harder than 200 eV:  [\ion{Fe}{11}] is not present, and most
  probably [\ion{Fe}{10}] neither, which will fit  with the lack of
  [\ion{Si}{7}] either (Fig.~\ref{fig9}).  \label{fig4}}
\end{figure}

 NGC~3227 shows the weakest coronal line region in the sample, and it is unresolved
at the spatial resolution achieved.
Figure~\ref{fig4} exhibits weak [\ion{Fe}{7}] and perhaps [\ion{Fe}{10}], the
latter being strongly blended with [\ion{O}{1}]~6363~\AA.
[\ion{Fe}{11}]~7889~\AA\ is not detected.
The coronal region is  restricted to the nuclear window, i.e., within 
 the inner 50~pc radius from the center. 
[\ion{O}{1}]~6300~\AA\ extends along the full length of the slit,
i.e., $\sim$300~pc radius in both north and south direction (only the
inner 130~pc radius is shown in Figure~\ref{fig4}).

[\ion{Si}{6}]~19630~\AA\  extends up to 45~pc north of the nucleus,
 consistent with the [\ion{Fe}{7}] size.
 [Si\,{\sc vii}]~24830~\AA\ is not detected, consistent with the non
 detection of  [\ion{Fe}{11}].  The absence of these high ionization lines
 indicates that a much softer ionizing continuum illuminates the 
NLR, lacking photons with energies
 $\geq$200~eV, than in the other galaxies of the sample.

For assessing the kinematics, [\ion{Fe}{7}] is the best isolated and higher 
S/N line. It is nevertheless affected by telluric residuals on its red side. 
A Gaussian fit to the line indicates a FWHM $\simeq 900$ \kms, almost
a factor 2 larger than that of [\ion{O}{1}]~6300~\AA\ (Table~\ref{fwhmopt}), and a  
centroid blueshifting of -125~\kms.

\subsubsection{NGC~3783 and MCG-6-30-15}

These two galaxies are the more distant and the only Type~1 Seyfert of
the sample.  The spatial resolution achieved was about 100~pc for
MCG6-30-15, $\sim$ 200~pc for NGC 3783. The actual 
spatial bin used in these cases is 20-50\% larger than the spatial resolution
(Table~\ref{log}), in order to prevent off-nuclear spectra get contaminated by
nuclear light. The nuclear windows are set by summing up all the spectra containing 
broad H$\alpha$ emission.  No NIR spectra are available for these
objects.

\begin{figure}
\epsscale{1.1}
\plotone{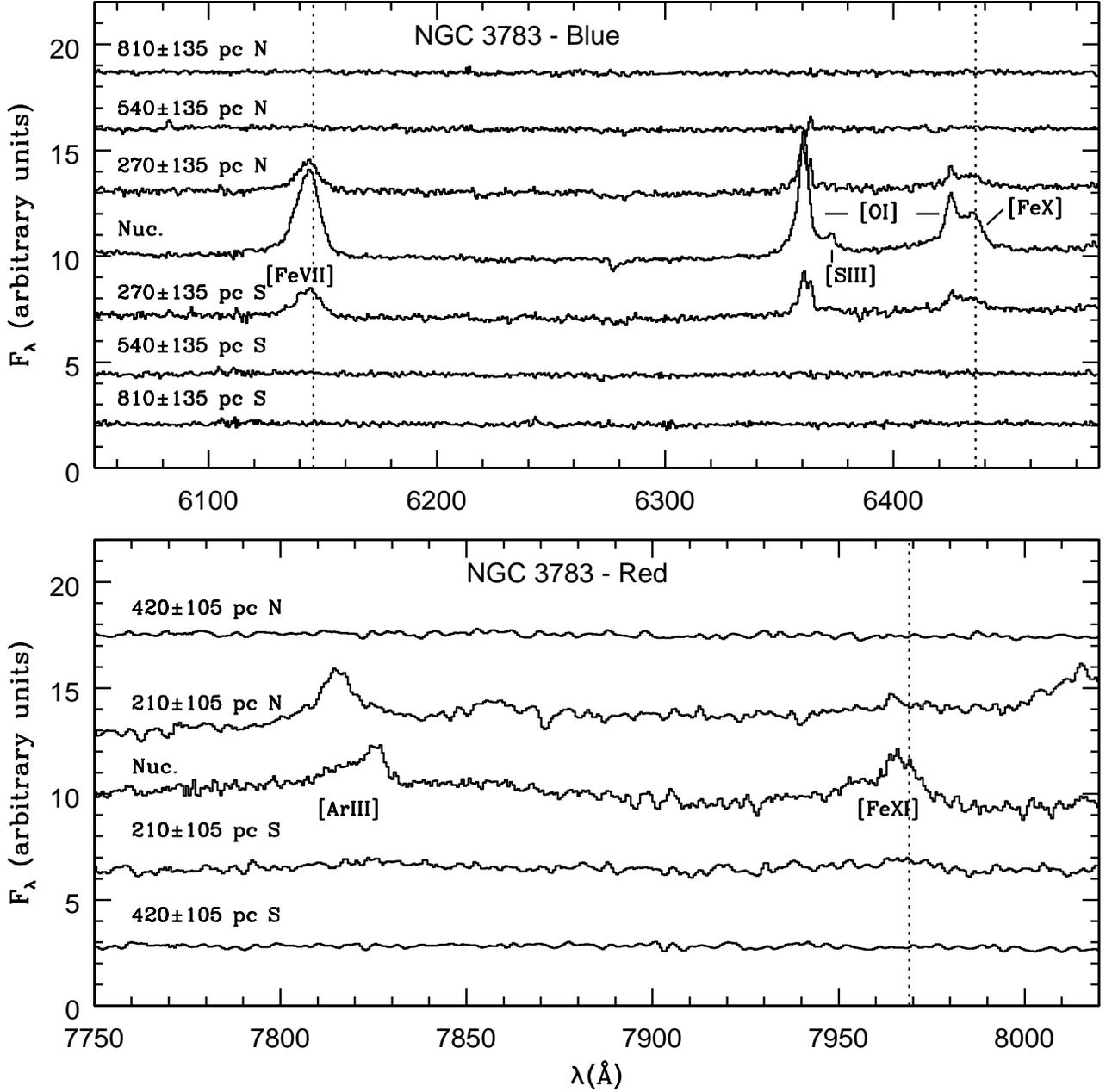}
\caption{The same as Fig~\ref{fig4} for NGC~3783. As in NGC 3227, the relative
  large size of the nuclear
  aperture window was dictated by  the number of spectra affected by
  scatter light from the  broad line region  - traced by H$\alpha$ in our
  spectra. This is to warrant the real spatial extension of 
coronal emission. \label{fig5}}
\end{figure}

{\it NGC\,3783} (Figure~\ref{fig5}) presents the largest coronal line region,
with more than 400~pc radius in the [\ion{Fe}{7}]~6087~\AA\ line - this is also
confirmed by [\ion{Fe}{7}]~5721~\AA\ line, present in all our spectra but not
shown for sake of simplicity.  [\ion{Fe}{10}] and 
[\ion{Fe}{11}] are also extended but to about half radius (Table~\ref{sizeclr}).
[\ion{O}{3}]~5007~\AA, however, is distributed in a halo around the
nucleus, extending up to a radius of 140 -- 175~pc (HST image in
\citet{sch03}).  This is about a factor 2 smaller than the [\ion{Fe}{7}]
region. We suspect that an over subtraction of the continuum image is
removing [\ion{O}{3}] emission from the outer regions of the halo. We note
that e.g., [\ion{Ar}{3}], (Figure~\ref{fig5}), is already very prominent at
210~pc from the nucleus.

The iron coronal lines are broader than [\ion{O}{1}] by about a factor of 2. At
the nucleus, they show a broad-blueshifted wing with FWHM $\sim$1000
\kms. At all spatial locations, the coronal peak position is
systematically shifted to the blue by more than 100~\kms\ (Table~\ref{fwhmopt}).
[\ion{O}{1}]~6300~\AA\ shows a clear double-peak profile, but [\ion{Fe}{7}]
does not: coronal gas is clearly more turbulent than neutral gas,
a double peak in [\ion{Fe}{7}] may be masked by the large width of the 
line.

\begin{figure}
\epsscale{1.1}
\plotone{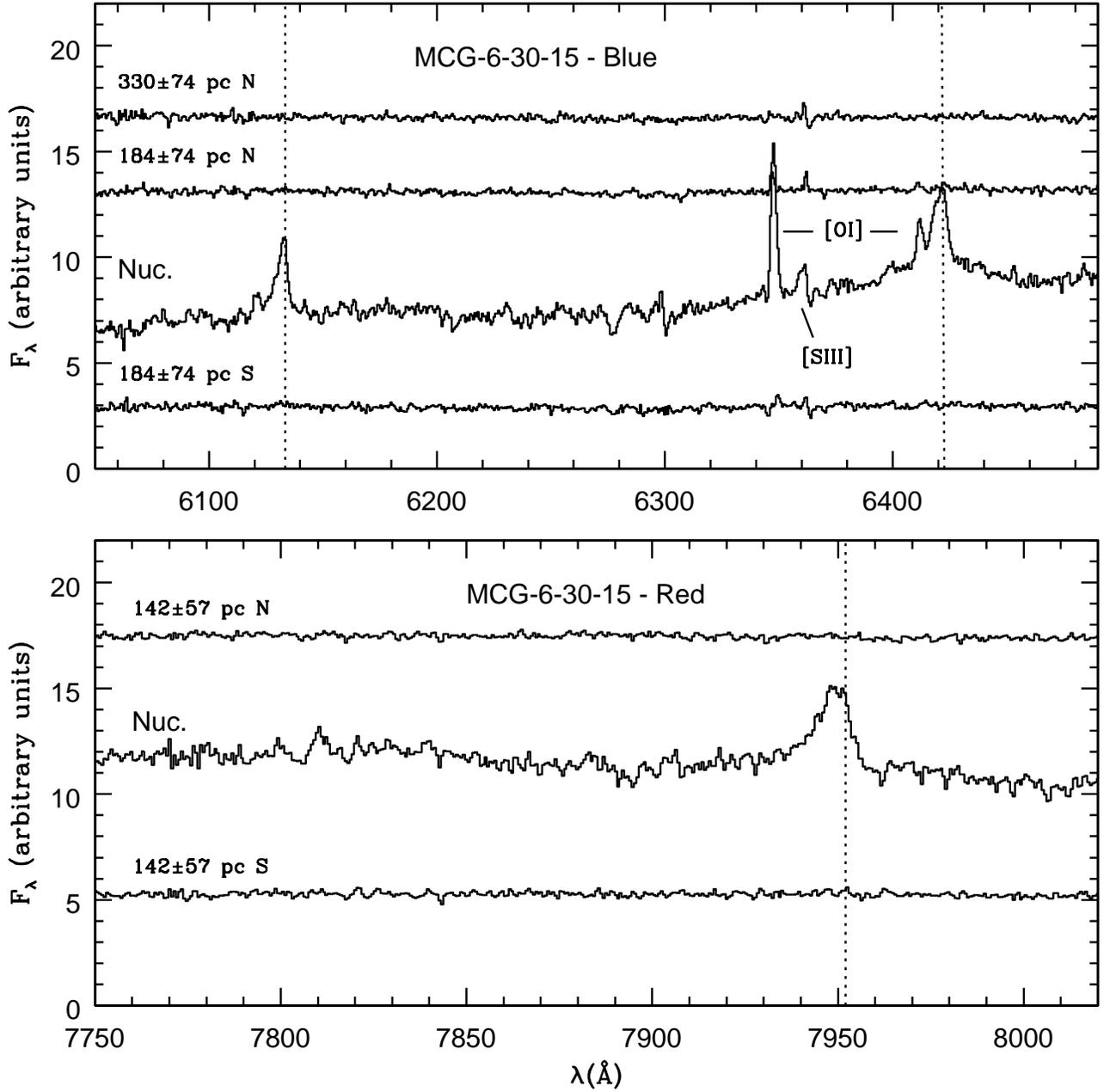}
\caption{The same as Fig~\ref{fig4} for MCG-6-30-15.  \label{fig6}}
\end{figure}

\begin{figure}
\epsscale{1.1}
\plotone{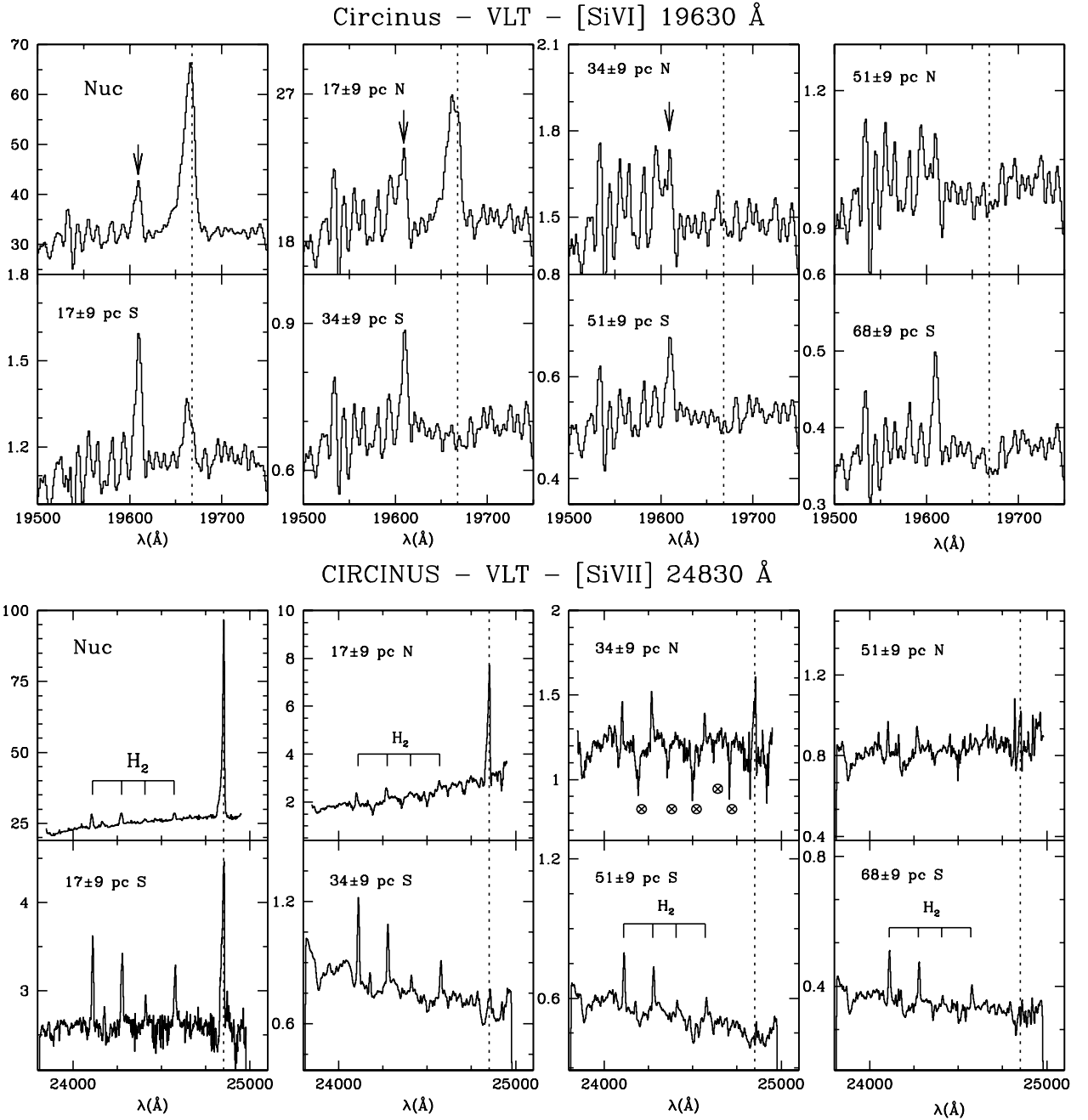}
\caption{NIR VLT spectra of Circinus. Upper and lower panels show the
spatial variation in the region around 
[Si\,{\sc vi}]~1.693~$\mu$m and [Si\,{\sc vii}]~2.47~$\mu$m, respectively. In the
upper panels the arrow indicate the expected position of H$_{2}$~1.957$\mu$m. 
Expected positions of the Q(0)H$_{2}$ lines are marked in the bottom panels.
The dotted line marks the position of the silicon line at the systemic
velocity. The orientation
of the slit is  north-south. The Y-axis is in units of erg~cm$^{-2}$~s$^{-1}$~\AA$^{-1}$.
\label{fig7}}
\end{figure}

\begin{figure}
\epsscale{1.1}
\plotone{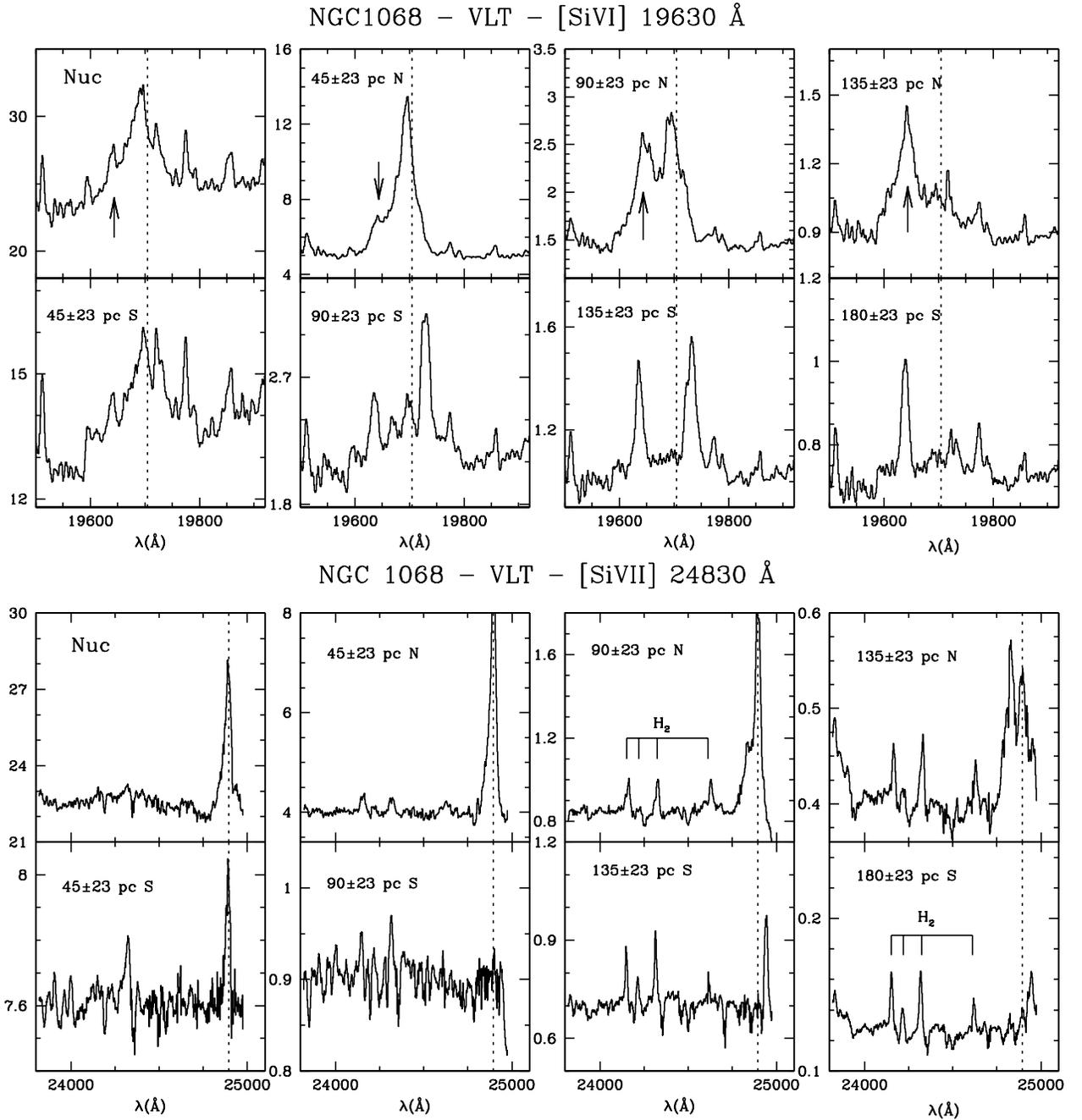}
\caption{VLT/ISAAC spectra for NGC~1068. The notation is the same as
in Figure~\ref{fig7}.   In NGC~1068
it is easily seen how the coronal gas moves faster than the more relaxed
molecular gas: this can be appreciated by comparing, in particular, the shift of the
[\ion{Si}{6}] line relative to that of the $H_2$ 1.957$\mu$m. \label{fig8}}
\end{figure}

\begin{figure}
\epsscale{1.1}
\plotone{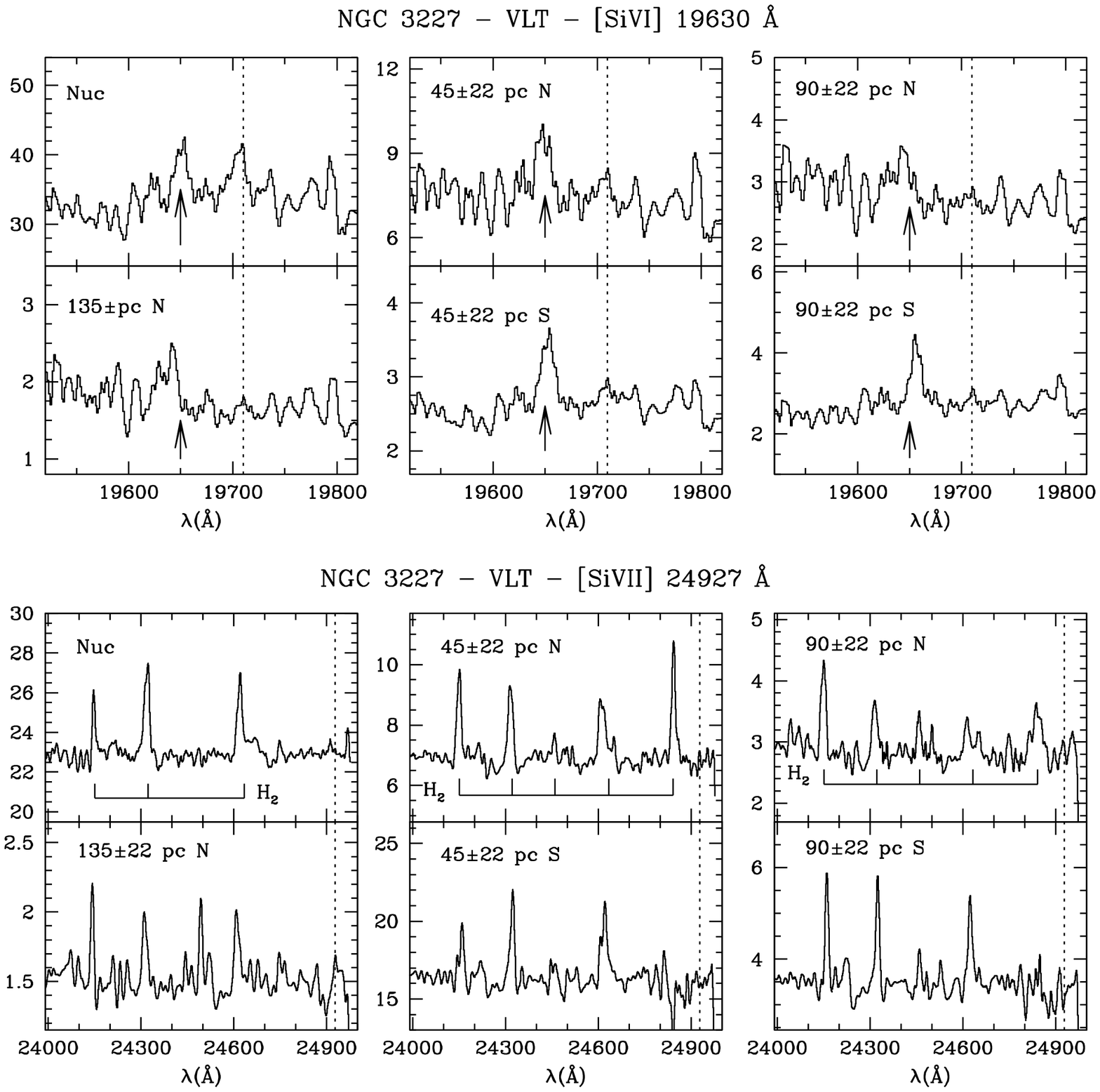}
\caption{VLT/ISAAC spectra for NGC~3227. The notation is the same as
  in Figure~\ref{fig7}. This galaxy lacks of [\ion{Si}{7}], which, as
  commented in Fig.~\ref{fig4} regarding to the absence of [\ion{Fe}{10}] and 
[\ion{Fe}{11}] in this object, supports the hypothesis of
  a very soft ionizing radiation continuum. \label{fig9}}
\end{figure}

\begin{figure}
\epsscale{1}
\plotone{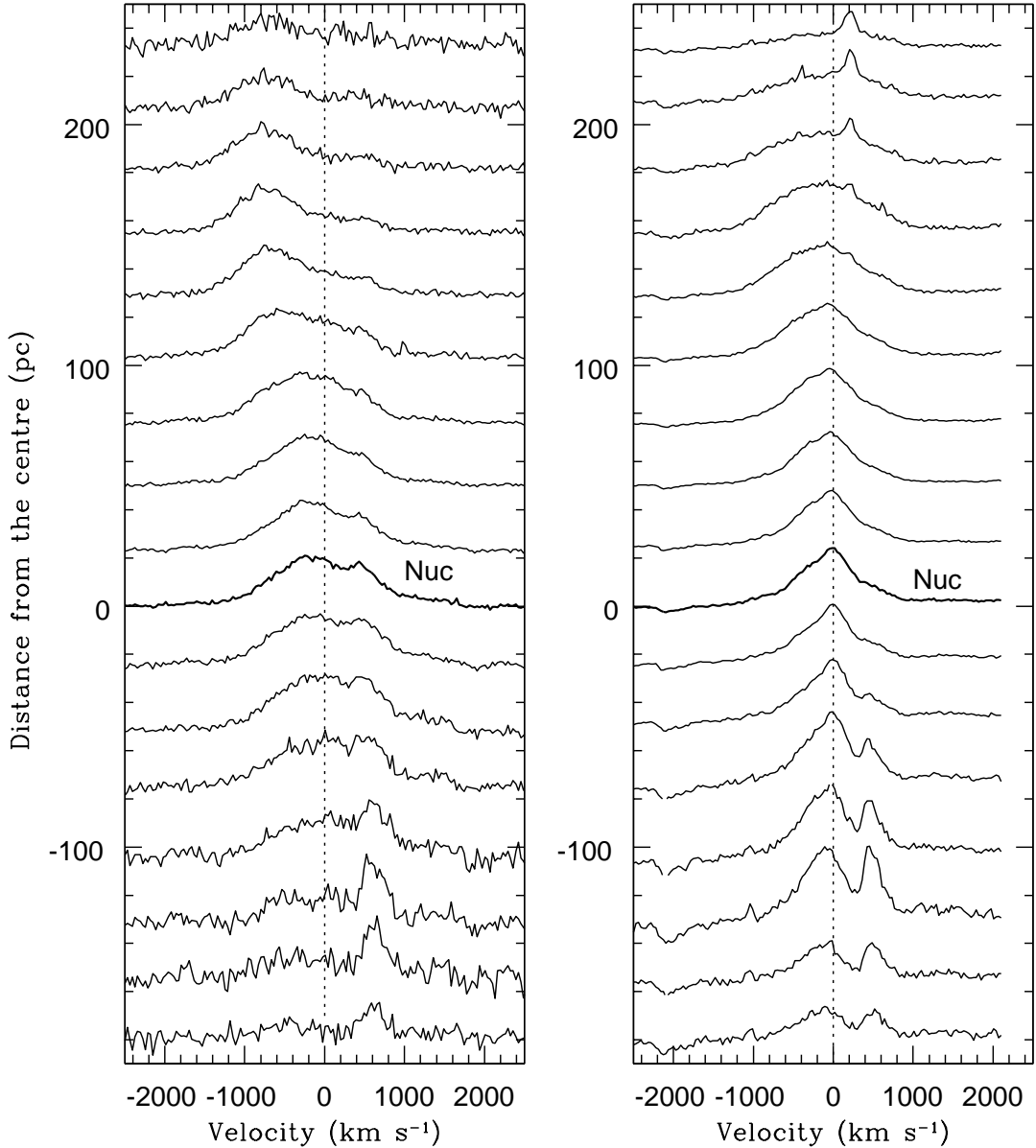}
\caption{Pixel to pixel variation along the spatial direction of the emission line profiles of [\ion{Fe}{7}]~6087~\AA\ (left panels) and [\ion{O}{1}]~6300~\AA\ (right panels), observed in NGC~1068. The
  dotted lines mark the position of the systemic velocity of the galaxy. The spectrum located at a distance of zero corresponds to the
  pixel containing the peak of light distribution. North is up and south is down.  The spectra are spaced by 26~pc from each other and are plotted in an arbitrary intensity
  scale. [\ion{Fe}{7}] shows a double peak component, each peak moving blueward and
redward from the systemic velocity at different spatial locations. We
envisage a geometry by which this gas is moving in radial directions
within a nuclear collimated wind, whose axis should be rather close to
the plane of the sky. By comparison, [\ion{O}{1}] is a slower gas. Its kinematics,
nevertheless, is more difficult to interpret because of the satellite line 
[\ion{S}{3}]~6312~\AA, seen south of the nucleus, and
presumably a [\ion{O}{1}] narrow component seen to the north, also seen in
[\ion{O}{1}]~6363~\AA\ in Fig.~\ref{fig3}. \label{fig10}}
\end{figure}

In {\it MCG-6-30-15} (Fig.~\ref{fig6}), the Fe coronal line emission is
  unresolved and thus restricted to the inner 100~pc radius in [\ion{Fe}{7}] 
  and [\ion{Fe}{10}] and less than 50~pc radius in [\ion{Fe}{11}]. In contrast,
  [\ion{O}{1}]~6300~\AA\ and [\ion{S}{3}]~6312~\AA\ extend up to
  300~pc north and south from the centre (this emission is not easily
  seen in Fig.~\ref{fig6} but cuts were chosen not to overcrowd the
  figure).  These low-ionization lines are also narrow,
  FWHM$\sim$ 100\kms, and symmetric whereas the Fe coronal lines show
  the typical blue wing, with FWHM $\sim$1500 \kms\ (Table~\ref{fwhmopt}). Both,
  the narrow and the broader component of the Fe lines, show a
  systematic blueshifting; that of the
  narrow component increases with IP, up to -140 \kms\ in [\ion{Fe}{11}]; the
  shift of the broad component is more uniform with IP, but larger,  in
  the range of $\sim -300$~\kms.

\section{The size of the  coronal line region} \label{comp}

Table~\ref{sizeclr} provides a compilation of the sizes of the coronal line
regions derived in this work. The table includes also coronal sizes
derived from the high spatial resolution 
[\ion{Si}{7}]~2.48~$\mu$m imaging of \citet{pri05}.  The smallest ``resolved'' coronal region is
found in Circinus, with a radius of $\sim$20 pc; the largest is found
in NGC 3783, with radius of 400~pc. Overall, the coronal region size
average in the 100~pc range. At each individual object, the CLR
size decreases with increasing ionization potential, i.e., the
hardest the photons the closer to the nucleus the highly ionized
ions are created.

Lower ionization lines  extend to much larger radius in
Seyfert galaxies, from several hundred pc to kpc.  Table~\ref{sizeclr} also
includes the sizes of [\ion{O}{1}]~6300~\AA\ and [\ion{S}{3}]~6312~\AA\
line regions extracted from the present data, and that of 
[\ion{O}{3}]~5007~\AA\ taken from the literature.  The coronal
line region is at least a factor 2--3 smaller than that of [\ion{O}{3}] or 
[\ion{O}{1}].  The most extreme case is Circinus, with a coronal
region a factor 10 smaller than that of [\ion{O}{3}].

These differences in sizes indicate a very stratified NLR, namely, 
lines with IP larger than 100~eV are restricted to the
inner 100\,pc radius whereas medium ionization gas, i.e, IP $<$54 eV,
extends to a region at least twice that value. Lines with IP $<30$ eV,
extend over a region comparable or larger than that of the medium
ionization lines. Photoionization produces this type of stratification,
and thus it should be  the principal source of gas
excitation in these galaxies.

Besides stratification, the ionized gas in Seyfert galaxies is known
to be collimated in large nuclear bicones seen in H$\alpha$ or [\ion{O}{3}]. We
know from previous work based on long-slit spectroscopy carried
out in directions both along and perpendicular to the ionization cone 
\citep{reu03}, and from direct imaging in the [\ion{Si}{7}]~2.48~$\mu$m line 
\citep{pri05}, that the coronal emission tends to extend 
preferentially along or within the ionization  cone traced by medium-ionization 
gas. Indeed, the latter work shows that the coronal gas is more collimated
than the medium ionization gas.
 
\section{Kinematics of the coronal gas} \label{kin}

Coronal lines are, in general, identified with a broad and blueshifted profile.
Assuming that the emission comes from an unresolved nuclear region, the 
systematic blueshifting has been  interpreted as  an 
outflowing wind provided that nuclear obscuring
material prevent us for seen the receding flow 
\citep[e.g.,][]{pe84,ev88,er97,rod02}.

\begin{figure}
\epsscale{1}
\plotone{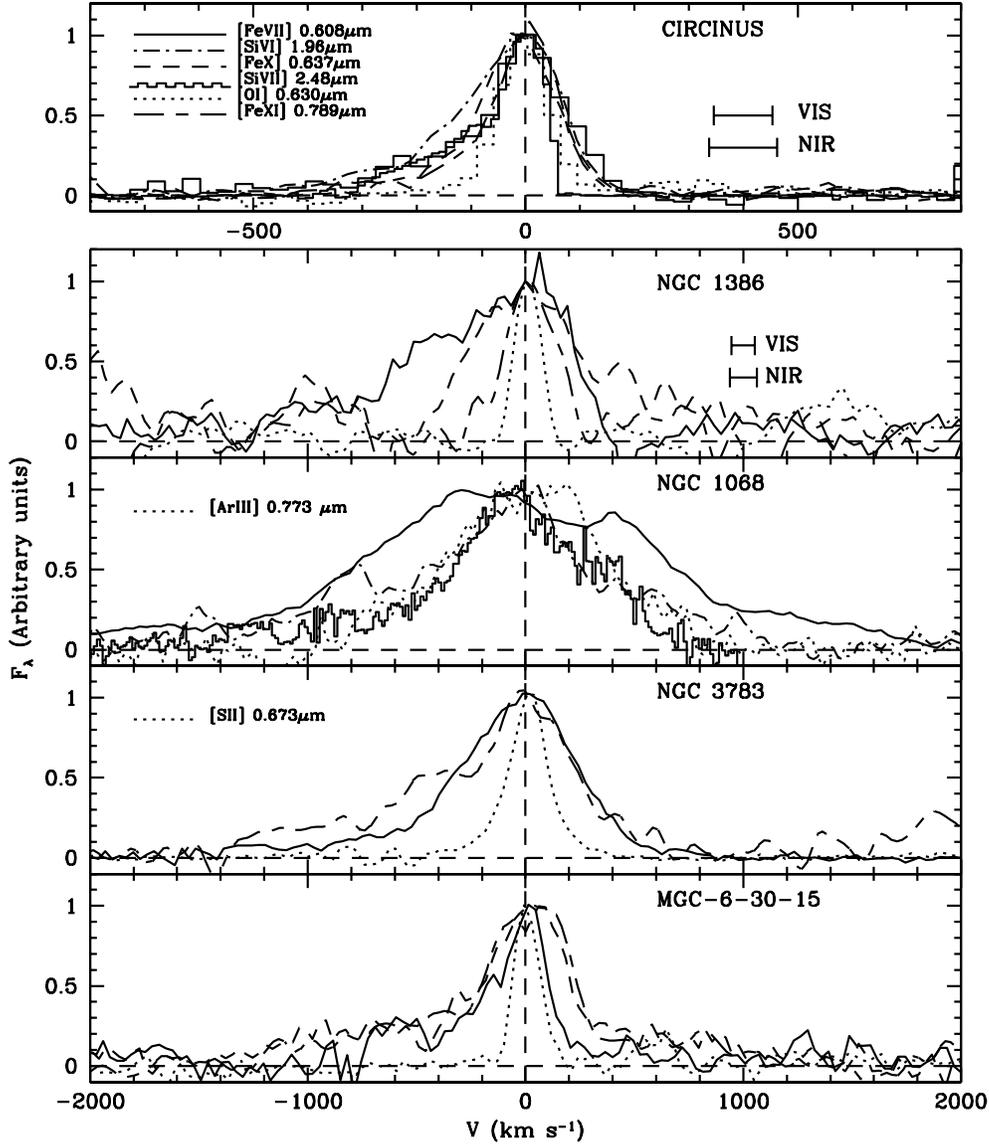}
\caption{Comparison of the nuclear line profiles in velocity space
for Circinus, NGC~1386, NGC~1068, NGC~3783 and MGC-6-30-15. Note that the velocity
scale in Circinus is different from that  in the other objects. 
In all cases, the low-ionization line profile (dotted line) 
is narrow - except in NGC 1068 - and symmetric
whereas the  coronal lines are significatively broader and asymmetric towards
the blue. The error bars represent the FWHM of the instrumental profiles,
 measured from the arc lamp (optical)  and sky-lines (NIR) respectively.
\label{fig11}}
\end{figure}

The present spectroscopic data conforms with this prototype  coronal profile  
albeit with some added complexity.  This can be seen in
Figure~\ref{fig11} , where the ``nuclear'' profiles  
of the best defined coronal lines for the objects
in the sample are compared.  For 
reference purposes, the profile of a
lower-ionization line is also included in the Figure. In all
objects a velocity-shift was applied to the
centroid positions to bring them all to V=0~km~s$^{-1}$.
Figure~\ref{fig11} shows that all coronal lines are asymmetric and
broader than the reference, low-ionization line. In most
cases, the profiles display a prominent blue asymmetry. At 20\% of
peak intensity, the blue wing extends up to 1200~\kms\ while the red
one goes only up to 600~\kms.  The less extreme case is Circinus: at
20\% of maximum, the blue and red wings extend only to 250~\kms\ and
100~\kms, respectively. The most extreme one is NGC~1068, with blue
and red wings extending in velocities up to 2000 \kms; the FWZI is larger
than 4000~\kms.  NGC~1068 and 
NGC~1386 are the only two objects displaying a double peak coronal profile, 
easily seen in [\ion{Fe}{7}] at both the nuclear and extended 
regions. To our knowledge, no previous detection of this such type of
coronal line profile has been reported for any other Seyfert.

The points below summarize the main  kinematic results:

1. In cases of high signal-to-noise and not blending, coronal line
profiles could be fit with a two Gaussian component: a narrow one with
FWHM comparable to, or slightly larger than, that of the lower
ionization lines, and a broad, usually blueshifted, component with FWHM ranging
from 700~\kms\ (e.g Circinus) to 1500~\kms\ (e.g MCG-6-30-15 or NGC
3783) and centroid shifts in the range 100--600~\kms.  The
centroid position of the narrow component remains in general at the
systemic velocity, as it is the case for the neutral gas lines,
[\ion{O}{1}] in the optical, H$_2$ in the near-infrared.

2. Blueshifting of the coronal emission is also seen in the spatial
resolved emission, and at both sides of the nuclear region along the
slit position (e.g., Circinus, NGC~1068 and NGC 3783).  If the emission is a
rather collimated wind, blueshifting at both sides of the nucleus would
require the axis of the wind to be rather close to the plane of the
sky in Type~2 objects, or close to the line of sight in Type~1. As the
counterpart receding gas is usually not seen, this   requires
the presence of obscuring material at the  base of
the wind.

3. One would expect an increase of the coronal line width with
   ionization potential, as the higher the ionization potential the
   closer to the ionization source the line is formed. Gas velocities
   should then approach those of the broad line region. 
   No such trend is generally seen  in the data.  As Fig.~\ref{fig11}
   illustrates, the Seyfert 1s MCG-6-30-15 and NGC~3783 show an
   increase in line width with ionization potential whereas in the
   Type~2 sources of the sample, the opposite is found. However this
   could be an obscuration effect in Type~2
   objects, which will affect mostly the higher ionization lines
   because they are formed closer to the center.
 
4. The kinematics of NGC~1068 deserves special attention. The best
 coronal emission kinematics is traced by [\ion{Fe}{7}].  The line profile
 both north and south of the nucleus shows a double peak component, visible
up to $\sim$250~pc and $\sim$180~pc, respectively, each
 tracing blueshifted and redshifted gas.  This type of profile has not ever
been reported in the literature for a Seyfert galaxy. 
One could think of several geometries to account for this complex
 kinematics, e.g., gas moving in circular orbits in a thick toroid with the
rotation axis close to the north-south direction, leading to a double
 peak spectrum at both sides of the nucleus. Alternatively, radial motions
 within a cone with its axis closely oriented to the plane of the sky 
 could lead to a double peak spectrum at different spatial locations 
 corresponding to approaching and receding gas. Observational 
 evidence found from polarimetry and the correlation between the 
NLR emission maps and the radio structure supports the later scenario. 

5. For the three galaxies for which optical and NIR spectra are
 available: Circinus, NGC~1068, and NGC~3727, the comparison of the
 respective coronal spectra indicates little extinction in the coronal
 region. This is based on the following arguments. First, coronal lines of
 comparable ionization potential are equally present in the optical as
 in the NIR, e.g. [\ion{Fe}{7}] and [\ion{Si}{6}]. Second, the coronal 
 region extends to
 similar sizes in the optical as in near-IR. Third, in Circinus, 
 optical and NIR coronal lines  present comparable  line profile. 
 Considering that the extinction is
 reduced by about 6 magnitudes between these two spectral regions,
 differences in the dispersion velocities due to differences
 in optical depth should be expected. It follows that optical coronal
 lines should be narrower than their NIR counterparts, which is not
 the case.

\section{Is photoionization driving the coronal line emission?} \label{mod}

Coronal lines are collisionally excited and can be produced either by
a gaseous region photoionized by a hard UV continuum \citep{fer97}, 
a hot ionized plasma \citep{vac89} or a mix of both \citep{con98}. 
Radiation from the central source alone cannot explain, however, the
systematic blueshifting of the lines. An alternative approach still
invoking photoionization was suggested by \citet{bin98}, who
proposed that radiation pressure from the central engine generates a
strong density gradient within the photoionized structure and is
responsible for accelerating the matter-bounded gas (the zone of high
excitation), explaining the systematic blueshift observed for coronal
lines. 

Moreover, constraints set by the observations regarding the nature of the
coronal regions are: the size of the emitting clouds are expected to
be less than 2~pc in Circinus and less than 7~pc in NGC~1068 -- on the basis
of the spatial resolution achieved in the [\ion{Si}{7}] images by \citet{pri05}. 
Cloud velocities, if reflecting any NLR velocity
field of accelerating gas, should reach up to 1000~\kms.

In order to investigate the most plausible scenario for the production
of the coronal lines taking into account the above constrains, we first 
tested the possibility that the coronal region, which
extends up to several tens of parsecs from the central engine, can be
produced by pure photoionization. To this purpose, models with the AANGABA code
\citep{gru97} were run.  The ionizing spectrum
suggested by Oliva et al. for Circinus (1999, see their Fig. 7) was
adopted; solar values \citep{ga89} were assumed for the
gas chemical abundances.

\begin{figure}
\epsscale{0.8}
\plotone{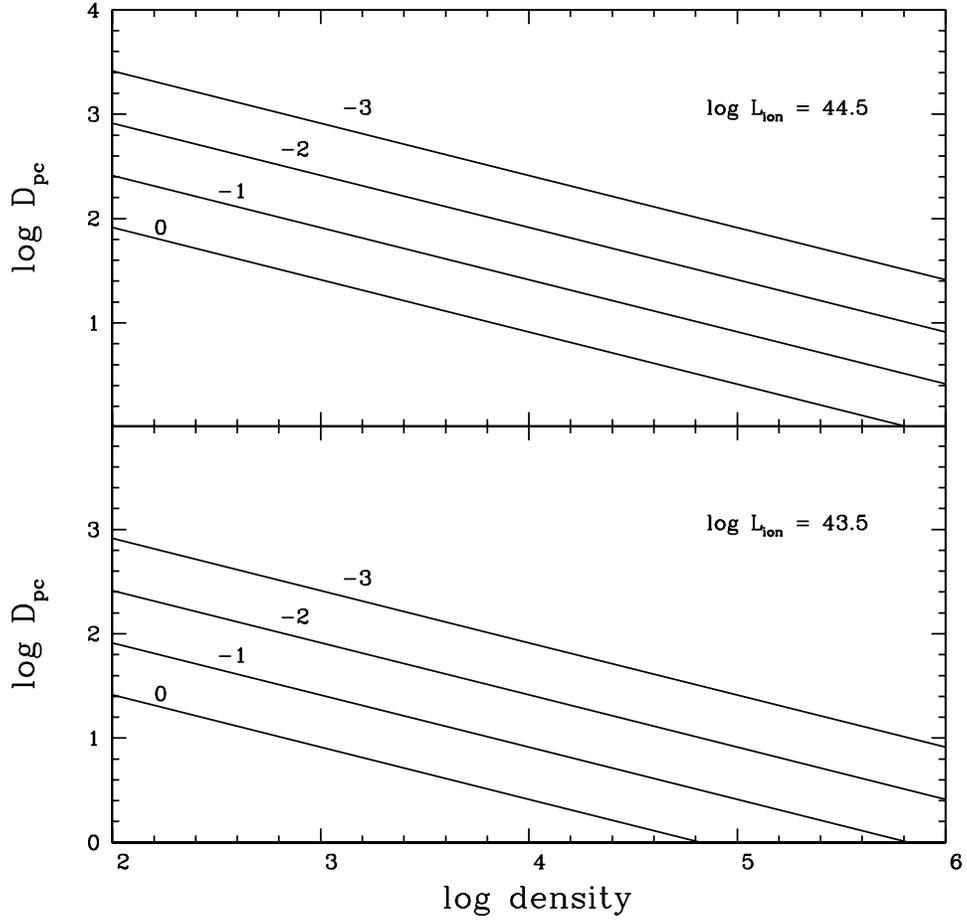}
\caption{Relationship between D, the distance from the ionizing source
(in parsecs) to the inner face of the cloud, and the gas density (in
cm$^{-3}$) for $L_{\rm ion} = 10^{44.5}$ erg s$^{-1}$ (upper panel)
and $L_{\rm ion} = 10^{43.5}$ erg s$^{-1}$ (lower panel). 
The curves are labelled according to the corresponding 
logarithm of $U$. \label{fig12}}
\end{figure}

Two values for the luminosity of the ionizing radiation were adopted:
$L_{\rm ion}$ = 10$^{43.5}$ erg s$^{-1}$ and 10$^{44.5}$ erg s$^{-1}$, as
used e.g. by \citet{fer97} and \citet{ol99} in their
respective photoionization models.  The number of ionizing photons,
$Q_{\rm H}$, contained by those two spectral energy distributions are
2.5$\times$ 10$^{53}$  and 2.5$\times$ 10$^{54}$ phot~s$^{-1}$, 
respectively.  A
grid of models was built with the density in the range 10$^2$ \cm3 to
10$^6$ \cm3, and the ionization parameter {\it U} = $\Phi_H$/{\it nc}
from 10$^{-3}$ to 1, where $\Phi_H$ is the flux of ionizing photons
reaching the coronal region, defined as $\Phi_H$ = Q$_{\rm
H}$/4$\pi$D$^2$, {\it D} being the distance from of the emitting cloud to
the central source, assuming the covering factor unity.  For each
$\Phi_H$ - or equivalently $L_{\rm ion}$ - there is a relation
between {\it U, n} and {\it D} \citep[see, for example,][]{fer97}.  
This means that for a given AGN, i.e., a given $L_{ion}$, {\it
U} is not a free parameter anymore if we know {\it D} from the
spatially resolved observations. The {\it U}, {\it n} and {\it D}
relation is given in Figure~\ref{fig12}.  Since coronal lines require
high ionization parameters \citep{fer97}, the distance from
the nucleus to the CL emitting cloud is limited, as we shall see 
below.

\begin{figure}
\includegraphics[angle=90,scale=0.8]{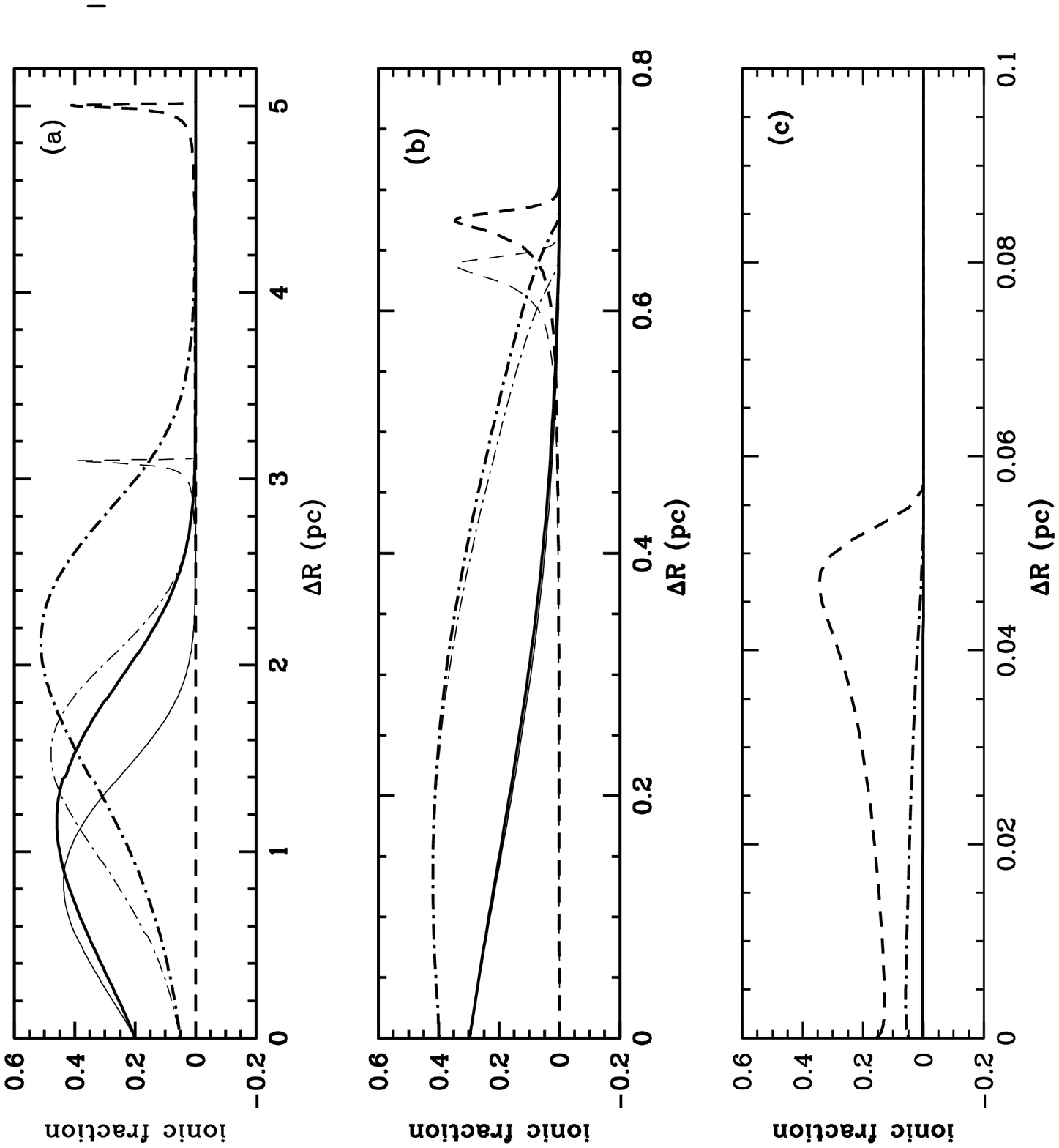}
\caption{Ionic abundance for the Fe ions versus the geometrical depth
of the cloud $\Delta R$, for $L_{\rm ion} = 10^{43.5}$ erg s$^{-1}$ (light 
lines) and $L_{\rm ion} = 10^{44.5}$ erg s$^{-1}$ (thick lines): 
Fe$^{+10}$ (solid lines), Fe$^{+9}$
(dashed-dotted lines) and Fe$^{+6}$ (dashed lines).  Panels a, b, and c show
the results for models with $U=$1, 0.1 and 0.01 respectively, and
density, n$_{\rm e}$ = 10$^4$~cm$^{-3}$. In panel (c), the results corresponding
to the two values of $L_{\rm ion}$ coincide and Fe$^{+10}$ is not present in 
these clouds}
\label{fig13}
\end{figure}

\begin{figure}
\includegraphics[angle=90,scale=0.8]{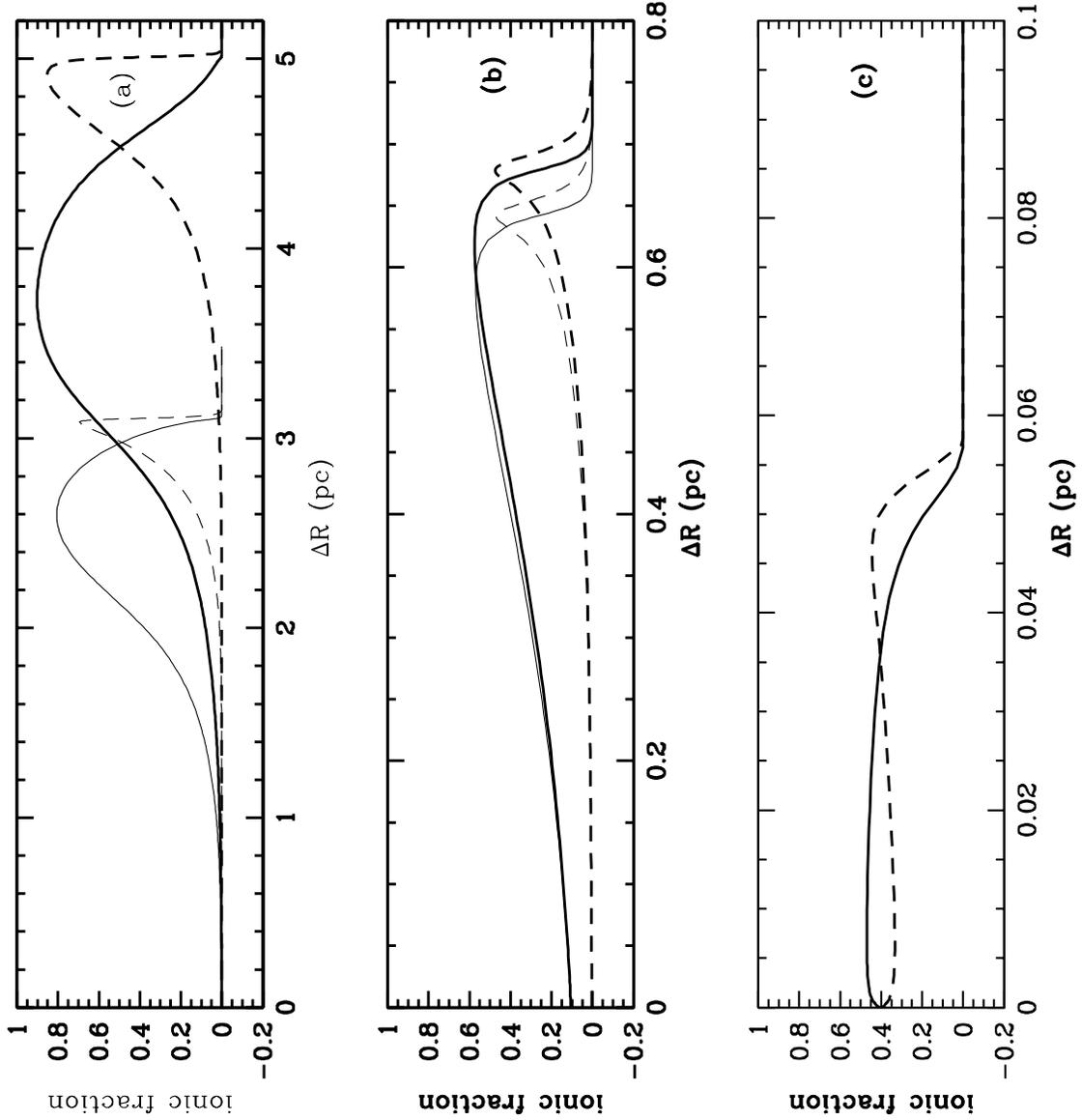}
\caption{Ionic abundance of the Si ions versus the geometrical depth
of the cloud $\Delta R$, for $L_{\rm ion} = 10^{43.5}$ 
erg s$^{-1}$ (light lines) and $L_{\rm ion} = 10^{44.5}$ 
erg s$^{-1}$ (thick lines): Si$^{+6}$ (solid lines), Si$^{+5}$ (dashed lines).  
Panels a, b, and c show the results for models with $U$=1, 0.1 and 0.01, respectively, 
and density, n$_{\rm e}$ =10$^4$~cm$^{-3}$. In panel (c), the results corresponding
to the two values of $L_{\rm ion}$ coincide.} \label{fig14}
\end{figure}

With the above parameters in mind, the distribution of the fraction of
iron and silicon ions producing the observed coronal lines versus the
size of the emitting cloud was computed. The results are shown in
Figure~\ref{fig13} and~\ref{fig14}, respectively, for $n_{\rm e}$ = 10$^4$
\cm3\ and $U$ = 10$^{-2}$, 10$^{-1}$ and 1. Ion fractions for models with 
$U$ = 10$^{-2}$ (panel c) and the two values of L$_{ion}$ coincide, and Fe$^{+10}$
is not present in these clouds. A density of 
$n_{\rm e}$ = 10$^4$ \cm3\ is
taken as face value following the estimate derived for Circinus from
[\ion{Ne}{5}]~14.3$\mu$m/24.3$\mu$m \citep{moor96}. Moving 
to higher densities will press more for  higher ionization
parameters, but in that case, following  Figure~\ref{fig12}, 
it will be difficult to explain the spatial extension of the coronal region.

It can be seen from Figs~\ref{fig12} and~\ref{fig13} that only the 
single-cloud models with $U \geq$ 0.1
show Fe$^{+6}$, Fe$^{+9}$ and Fe$^{+10}$.  Clouds with lower values of
$U$ tend to be dominated by Fe$^{+6}$ and less ionized species.  The
distributions of Si$^{+5}$ and Si$^{+6}$ are similar to that of
Fe$^{+6}$. The Figures also show that the geometrical
width of the clouds emitting the coronal lines is less than 1~pc for
$U$ = 0.1 and less than 5~pc for $U$ = 1, if $n_{\rm e}$ = 10$^4$ \cm3. 

Notice that the lower the density, the wider the coronal region size.  
Considering U$\leq$0.1 and an average coronal region size of 150~pc, it appears that a
density of $n_{\rm e}$ = 10$^4$ \cm3\ is too large to account for the observed
dimensions (Fig.~\ref{fig12}). We thus conclude that densities down to at least $n_{\rm e}$ = 10$^3$
cm$^{-3}$ are required.  Lower densities will lead to even larger
regions, but that will penalize line detection - as the emission is
proportional to $n_{\rm e} ^2$.  Larger distances could be reached
if the
filling factor were low. Yet,  very high spatial resolution images in
[\ion{Si}{7}]~2.48~$\mu$m of Seyfert galaxies 
\citep{pri05} show that the
coronal emission is not yet resolved in blobs/ knots
at  spatial scales of  10--15~pc,  but  shows
instead diffuse morphology. With such   low surface brightness 
it will be difficult to detect any emission if the filling factor were low.

Thus, assuming that lines from different ionization potential ions, e.g.,
[\ion{Fe}{7}], [\ion{Fe}{10}] and [\ion{Fe}{11}], {\it are all produced in
the same cloud}, a cloud density of $n_{\rm e}$= 10$^3$ \cm3, and $U$ between 0.1
and 1, according to Figure~\ref{fig12} the maximum size of the coronal 
line region is $\leq$100~pc for the brighter source
($L_{ion}$ = 10$^{44.5}$ erg~s$^{-1}$), and $\leq$30~pc for the 
dimmer one if it is powered entirely by photoionization.

\begin{figure}
\plotone{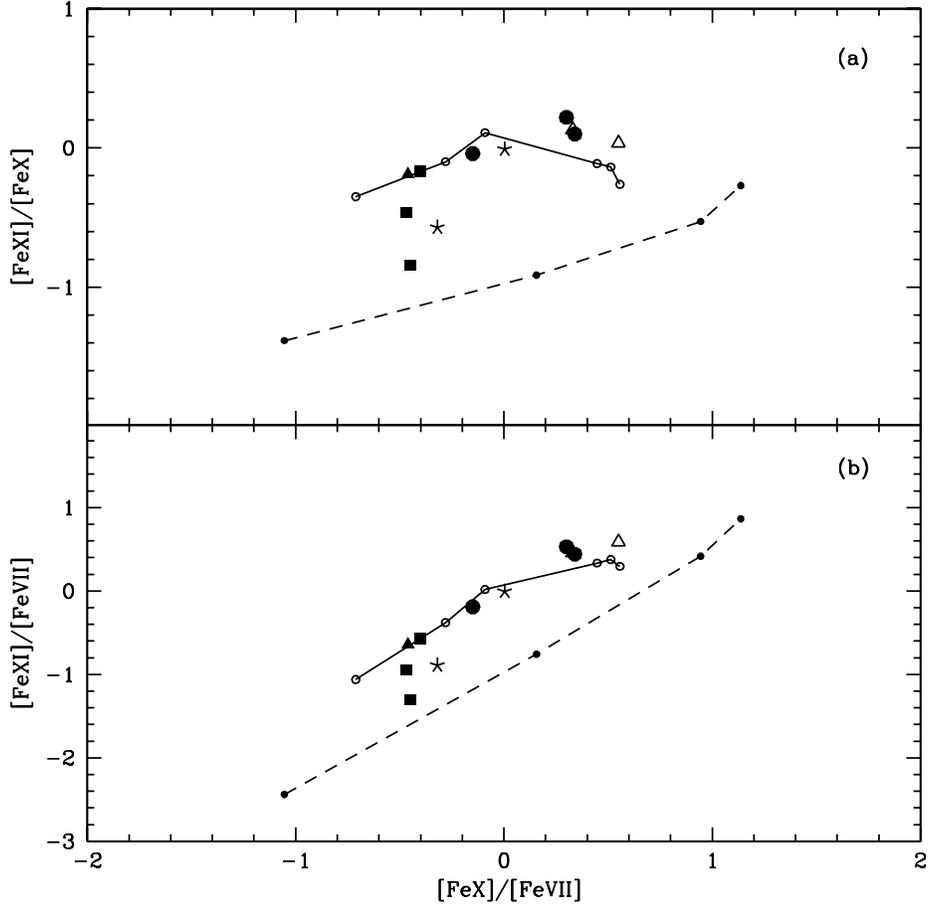}
\caption{Observed Fe coronal line ratios for the nuclear and extended region measured in the AGN sample compared with  predictions from pure
photoionization models (dashed line, dots correspond to different
values of U ( -2, -1.5, -1, 0) for $n_{\rm e}$ = 3$\times 10^3$ cm$^{-3}$ and L$_{ion}$ =
10$^{44.3}$~erg s$^{-1}$) and shock dominated models (solid line). For
the later, the shock velocity varies from 300~\kms\ to 900~\kms,
increasing from left to right. Most data points fall in the shock
dominated region.  Symbols correspond to different galaxies: filled
circles, Circinus; filled triangles, NGC~1386; filled squares,
NGC~1068; open triangles, MGC-6-30-15; and stars,
NGC~3783. \label{fig15}}
\end{figure}

Diagnostic diagrams based on flux ratios between the three optical 
iron lines provide a
good test to the photoionization models and are shown in 
Fig.~\ref{fig15}. Because of the relationship between $U$, $n$ and
$D$, the loci of the results from all the models are represented by
the same curve (dashed). On this curve, the dots correspond to
different values of $U$ (log $U$ = -2, -1.5, -1, and 0) 
for $n_{\rm e}$ = 3$\times 10^3$ cm$^{-3}$ and $L_{\rm ion}$
= 10$^{44.3}$~erg s$^{-1}$. Note that a density of 3$\times 10^3$ cm$^{-3}$
was chosen because it is located between the two limiting cases, i.e.,
$10^3$ and $10^4$ \cm3.
Data plotted in this Figure are extracted from
Table~\ref{fwhmopt}. The line ratios show an ample disagreement with these
standard photoionization models.

The only way to single-cloud photoionization models reproduce 
the iron line 
ratios is to find a way to shift the photoionization results up and left 
on both panels of Fig.~\ref{fig15}. This could be achieved  with an increase 
of the [\ion{Fe}{11}] line with respect to the [\ion{Fe}{7}] 
line, and the later relative to [\ion{Fe}{10}]; or the opposite: a decrease of  the [\ion{Fe}{10}] line with respect to [\ion{Fe}{7}] and [\ion{Fe}{11}]. 
Our results were obtained using effective collision
strengths calculated using the distorted wave method \citep{kc70,ns82}. 
More recently, atomic data for  those 
lines have been available in the literature,  mainly due to the so 
called  Iron Project \citep{hum93}. These calculations are 
carried out using the 
R-matrix method and account for resonances. These resonances
tend to increase the excitation cross-sections, and the net effect is 
an increase of the collision strengths, mainly at lower temperatures 
\citep{agg03,ber00,pel01}. A close look into the new data shows 
that indeed  the new collision strengths are  higher than the values 
used in this paper, as expected. Compared to the previous values used in
our calculations, the increase is more significant  
for \ion{Fe}{10} than for the lines of the other two ions, 
\ion{Fe}{11} being the second one more affected. Thus, 
the [\ion{Fe}{10}]/[\ion{Fe}{7}] ratio should increase if calculated 
with  new atomic data , as well as [\ion{Fe}{11}]/[\ion{Fe}{7}], 
although by a smaller factor. On the other hand, 
[\ion{Fe}{11}]/[\ion{Fe}{10}] should decrease. In brief, the overall 
tendency is to shift the photoionization results presented in 
Fig.~\ref{fig15} (dashed line) to the right, leading to a larger 
disagreement with the observational data.

We then consider the additional effect of shock excitation ``coupled'' with 
photoionization to explain the coronal emission.
As a constraint, we assumed that the observed velocity width (at FWHM)
measures the actual shock velocity that passes through the gas, which should 
then be in the range 300 to 900~\kms. The resulting composite 
models are also shown in
Fig.~\ref{fig15} (solid line). They were generated using the code SUMA,
which accounts for the coupled effect of shock and photoionization
\citep{vc97}. In these composite models $L_{\rm ion}$  has
the same  values as those used in the pure photoionization models.

If the coronal region clouds are shock dominated, there is no upper
limit for the distance of the coronal emission to the nucleus as there is
for the pure photoionization case. In order to limit the average
cloud density to 10$^4$ \cm3, models with shock velocities higher than
500~\kms\ were built assuming a preshock density n$_0$ = 200 \cm3,
while for lower velocities, $n_0$ = 300 \cm3\ was assumed. Models
shown in Figure~\ref{fig15} illustrate the contributive
effect of shocks to excite coronal lines in the shock-dominated case. 
A composite effect of shock and photoionization 
potentially offers then a better agreement of the observed iron line 
flux ratios. Notice that the effect of the newer collision strength data
is less important in the post-shock region, where the iron coronal lines are 
produced, because the temperature is higher than in the photoionized zone. 
A smaller shift of the shock-dominated curve  to the right would indicate
that velocities lower than 300~\kms\ should probably also be included.

The models presented provided a
first insight into the physical processes powering the
CLR. Self-consistent models for a given object would require a more
comprehensive analysis of the whole emission-line spectrum, which is out
of the main focus in this paper.  As
already discussed for some AGN \citep{con98,con98b,con03}, we expect that
photoionization as well as shock must contribute to the physical
conditions of the different types of emitting clouds present in the
NLR. The results shown in Figure~\ref{fig15}  support this
scenario.

\section{Concluding remarks} \label{con}

\begin{itemize}

\item Coronal lines are expected to be presented in all type of
 AGN. This work surveyed at unprecedented spectral and spatial resolution, 
five coronal lines covering a large
 range both in ionization potential, between 100 and 260 eV, and
 wavelength (optical and NIR regions) in six Seyfert
 galaxies: three Type~2, two Type~1 and one Type~1.5. 
 We found that they are all present in the galaxies with the exception of
NGC~3727, where only ions from the lowest ionization potentials are
detected.  Thus, the shape, or hardness, of the ionizing spectrum in
 Seyfert galaxies is very similar  regardless of the type. This
 is in the same line as found by \citep{pr00} in a different
 sample of Seyfert galaxies, after the
 analysis of IR-ISO coronal lines spanning a range in ionization
 potential between 54 and 300~eV. NGC~3727 is the only case  
showing very soft ionizing spectrum as only lines from the lowest IP are present.

\item The coronal region is spatially resolved over scales that range
 from a few tens of parsecs up to a few hundreds of parsecs from the
center. The size of the emitting region varies with  the
ionization potential: the highest the ionization potential the most
compact the  region becomes. This stratification indicates that nuclear
photoionization is the principal excitation mechanism.

 \item Coronal line profiles are characterized by two
components: a narrow one whose centroid and FWHM values are within
the range found for lower ionization lines, and a broader one,
 whose centroid is systematically shifted to the blue by a
few hundreds of \kms\ and has a FWHM  a factor 2 at 
least larger than that  of 
the lower ionization lines.  The gas velocities implied by this
component vary from object to object, and are in the range  from 500
up to 2000~\kms. This blueshifting is interpreted as an 
outflowing wind. However, a nuclear wind
should have an approaching and a receding component.
As the blueshifted component is the one often seen, the redshifted one 
has to be obscured by dust at the base of the wind.

\item We are however able to see the receding component of the wind in two
objects: NGC~1068 and NGC~1386. This interpretation comes from the
detection of a spatially 
resolved double peak in the [\ion{Fe}{7} line, each peak tracing respectively
blueshifted and redshifted gas at each spatial location. 
A spatially resolved double peak could be produced
by gas moving in circular orbits in a thick toroid with the
rotation axis close to the north-south direction, or by radial motions
along a collimated cone, with the axis closely oriented to the plane 
of the sky. The observational evidence supports this latter 
scenario.

\item For a luminosity of the ionizing radiation source in the
range 10$^{43.5}$ - 10$^{44.5}$ erg s$^{-1}$, single-cloud
photoionization models indicate that  to produce detectable coronal
emission over a wide range of ionization potentials,   the
ionization parameter must be larger than 0.1 and the density should be
larger than 10$^3$ cm$^{-3}$. Under these conditions, 
such simple single-cloud models can account
for coronal emission of up to 100~pc from the
nucleus if the ionizing radiation luminosity is higher 
than 10$^{44}$ erg~s$^{-1}$. For lower luminosities the size of 
the coronal regions can at most reach 50~pc, which is  smaller 
than the results obtained for some of the galaxies of our sample. 
In addition,  these models fail to explain the observed Fe line ratios. 
Furthermore, even  multi-cloud photoionization models, mimicking 
the stratification of the NLR, can hardly reproduce the  observations, 
leading us to conclude that another energy source must  be present.

\item In support to the above result, diagnostic diagrams using the 
Fe lines show
that single-cloud photoionization models largely depart from the
trend shown by the data. However, if the contribution of additional
shock excitation is included, with shock velocities between 300 and
900~\kms\ in agreement with observations, 
the {\it combined} photoionization plus shock effect is able to account for
the observed ratios.

\item It is interesting to remark that the systematic presence of H$_2$
lines in the nuclear region of AGN implies very large variations in
the physical conditions of the gas, capable of sustaining highly
ionized species along with molecules which easily dissociate with the
radiation of the central engine.  The molecular H$_{2}$ emission clearly relates to a
different type of clouds from that traced by the ionized gas, on the
basis of both its kinematic - low rotation, narrow line widths - and
spatial distribution which usually extends over several hundreds
parsecs to kpc from the nucleus, and preferentially in directions
perpendicular to the ionization cone \citep{reu02,reu03,rod04,rod05}.

\end{itemize}

\acknowledgments

This research has been partly supported by the Brazilian agencies
FAPESP (00/06695-0) to SV and RW, CNPq (304077/77-1) to
SV and CNPq (309054/03-6) to ARA. This research has made use of the 
NASA/IPAC Extragalactic Database (NED) which is operated by the Jet 
Propulsion Laboratory, California Institute of Technology, under 
contract with the National Aeronautics and Space Administration.

\clearpage

\begin{deluxetable}{lcccc}
\tablecaption{Properties of the Seyfert galaxies studied in this work. \label{pa}}
\tablewidth{0pt}
\tablehead{
\colhead{Name} & \colhead{Seyfert Type} &  \colhead{pc/$\arcsec$} & \colhead{z} &  

\colhead{PA ($^{\rm o}$)\tablenotemark{a}}}
\startdata

CIRCINUS & 2 & 19 & D$\sim$4 Mpc\tablenotemark{b}  & -44\arcdeg NW        \\    
NGC~1386  & 2 & 56 &  0.00289 &  0\arcdeg N-S      \\  
NGC~1068 & 2 &  75 & 0.00379  &  35\arcdeg NE     \\      
NGC~3227 & 1.5 & 74 & 0.0038   & 10\arcdeg NE     \\  
MCG-6-30-15  & 1 & 105 & 0.00775  & -65\arcdeg NW        \\  
NGC~3783 & 1 & 192 & 0.00973  &  -10\arcdeg NW  \\
\enddata
\tablenotetext{a}{Derived from [\ion{O}{3}]~5007~\AA\ line images.} 
\tablenotetext{b}{\citet{u3}}
\end{deluxetable}

\begin{deluxetable}{lcccccccc}
\tabletypesize{\scriptsize}
\tablecaption{Log of Observations. \label{log}}
\tablewidth{0pt}
\tablehead{
\colhead{Galaxy} & \colhead{Date of Obs.} &  \colhead{Telescope} & \colhead{Instrument} &  
\colhead{Seeing} & \colhead{Airmass} & \colhead{D}\tablenotemark{1} & \colhead{$\lambda_{\rm c}$} & \colhead{exp. time} \\
                 & \colhead{(dd/mm/year)}   &      &    &
\colhead{($\arcsec$)} &  & \colhead{($\arcsec$)} 
& \colhead{(\AA)} & \colhead{sec.} \\
\colhead{(1)}   & \colhead{(2)}   & \colhead{(3)}   & \colhead{(4)}   & \colhead{(5)}   &  \colhead{(6)}   &
\colhead{(7)}   & \colhead{(8)}   &\colhead{(9)}}  

\startdata

NGC~1068 & 31/01/2002  & ESO--NTT & EMMI  & 1.33  & 1.40 & 1.4  & 6230 & 4$\times$400  \\  
         & 10/03/2002  & ESO--NTT & EMMI  & 0.98  & 1.10 & 1.1  & 7890 & 4$\times$400 \\   
         & 24/11/2001  & ESO--VLT & ISAAC & 0.72  & 1.21 & 0.6  & 19800 & 10$\times$60 \\  
         & 24/11/2001  & ESO--VLT & ISAAC & 0.60  & 1.15 & 0.6  & 24400 & 10$\times$60 \\  
CIRCINUS & 28/02/2002  & ESO--NTT & EMMI  & 1.25  & 1.24 & 1.4  & 6230 & 4$\times$400 \\   
         & 28/02/2002  & ESO--NTT & EMMI  & 0.91  & 1.25 & 1.1  & 7890 & 4$\times$400 \\   
         & 06/04/2002  & ESO--VLT & ISAAC & 0.63  & 1.33 & 0.9  & 24400 & 10$\times$60 \\  
         & 06/04/2002  & ESO--VLT & ISAAC & 0.63  & 1.32 & 0.9  & 19800 & 10$\times$60 \\  
NGC~1386 & 27/11/2001  & ESO--NTT & EMMI  & 1.20  & 1.01 & 1.4  & 6230 & 4$\times$400 \\   
         & 27/11/2001  & ESO--NTT & EMMI  & 0.98  & 1.00 & 1.1  & 7890 & 4$\times$400 \\   
NGC~3227 & 28/02/2002  & ESO--NTT & EMMI  & 0.92  & 1.55 & 2.1\tablenotemark{a}  & 6230 & 4$\times$400 \\   
         & 28/02/2002  & ESO--NTT & EMMI  & 0.93  & 1.53 & 1.6 \tablenotemark{a} & 7890 & 4$\times$400 \\   
         & 10/02/2002  & ESO--VLT & ISAAC & 0.56  & 1.48 & 0.6  & 19800 & 10$\times$120 \\ 
         & 10/02/2002  & ESO--VLT & ISAAC & 0.56  & 1.48 & 0.6  & 24400 & 10$\times$120 \\ 
NGC~3783 & 10/03/2002  & ESO--NTT & EMMI  & 1.13  & 1.06 & 1.4  & 6230 & 4$\times$400 \\   
         & 10/03/2002  & ESO--NTT & EMMI  & 0.95  & 1.15 & 1.1  & 7890 & 4$\times$400 \\   
MCG-6-30-15 & 28/02/2002  & ESO--NTT & EMMI  & 1.05  & 1.08 & 2.1\tablenotemark{a}  & 6230 & 4$\times$400 \\
            & 28/02/2002  & ESO--NTT & EMMI  & 1.00  & 1.02 & 1.6\tablenotemark{a}  & 7890 & 4$\times$400 \\

\enddata
\tablenotetext{1}{D is the diameter of the extraction window used to
  define the nuclear spectrum and those of adjacent regions. Spectra
  shown in Fig. 1 to 9 are extracted according to this value.}

\tablenotetext{a}{Aperture size for the nuclear spectrum. Spectra for the extended regions were extracted using a window size of 1.4$\arcsec$ and 1.1$\arcsec$ for the blue and red
regions respectively}.  

\end{deluxetable}

\begin{deluxetable}{lcccccccc}
\tabletypesize{\scriptsize}
\tablecaption{Size  (radius in pc measured from the nucleus) of the  coronal and lower ionization line  regions  in  Seyfert galaxies\tablenotemark{1}. References are provided for
those values 
taken from the literature. \label{sizeclr}}
\tablewidth{0pt}
\tablehead{
\colhead{Object}  & \colhead{[Fe\,{\sc vii}]} & \colhead{[Fe\,{\sc x}]} & 
\colhead{[Fe\,{\sc xi}]} & \colhead{[Si\,{\sc vi}]} & \colhead{[Si\,{\sc vii}]} &
\colhead{[O\,{\sc i}]} & \colhead{[S\,{\sc iii}]} &  \colhead{[O\,{\sc iii}]} \\
\colhead{IP(eV)}&  \colhead{100} & \colhead{240} & \colhead{260} &
\colhead{170} & \colhead{205} & \colhead{0.0} & \colhead{24} & \colhead{35.1}}
\startdata
Circinus    & 53~N-27~S & 27~N-S & 21~N-S  & 17~N-S &  35~N-17~S &  200 & 27\tablenotemark{2} & $>$500\tablenotemark{a} \\
NGC~1386    & 108~N-49~S   & 49~N-S &  53~N-38~S &  NA  &  NA   &  235 & 157 & 165\tablenotemark{b}  \\
NGC~1068  & 210~N-105~S  & $<\sim$100 &  $<\sim$40 & 90~N-135~S & 135~N-S & 525 & 210 & 375\tablenotemark{b} \\
NGC~3227    & $<$78  &  $<$78 &  \nodata & 45~N &  \nodata  & 300 & 106 &  518\tablenotemark{c} \\
NGC~3783  & 437~N-302~S  & 235~N-302~S & 120~N &  NA  & NA  & 540~N-270~S & $<$135 & 175\tablenotemark{b} \\
MCG-6-30-15 & $<\sim$ 90 & $<\sim$90 & $<$50 & NA        & NA        &  $\sim$300 & $\sim$300 & 295\tablenotemark{b}\\
NGC~3081 & & & & &120~N\tablenotemark{d} & & & \\
ESO428-G014 & & & & &120~NW--160~SE\tablenotemark{d} & & & \\
\enddata
\tablenotetext{1}{N-S mean same distance to the North and South, NA = not
available. [O\,{\sc iii}]~5007~\AA\ sizes are measured from imaging data. [Si\,{\sc vii}]
images exist only for the four objects quoted in the table, two of
them are not included in this spectroscopy study, the other two,
Circinus and NGC~1068 show comparable size to that reported here from
spectroscopy.} 
\tablenotetext{2}{This value is one order of magnitude smaller than the one 
determined by \citet{ol99} using [S\,{\sc iii}]~9531~\AA. The
large discrepancy between the two sizes is likely due to the slit orientation, 
which in our case is along the edge of the ionization cone (PA=-7$\deg$) while in \citet{ol99} is along
its axis (PA=-44$\deg$). Clearly, the [S\,{\sc iii}]~9531~\AA\ emission is highly anisotropic and it
seems to follow that of [O\,{\sc iii}]~5007~\AA\ (see Figure~1 of \citet{ol99}).} 
\tablenotetext{a}{\citet{veb97}}
\tablenotetext{b}{\citet{sch03}} 
\tablenotetext{c}{\citet{mun95}}
\tablenotetext{d}{\citet{pri05}}

\end{deluxetable}

\begin{deluxetable}{lccccccccccccccc}
\tabletypesize{\scriptsize}
\rotate
\tablecaption{FWHM and shifts \tablenotemark{a} 
from the centroid position (both in km/s), and fluxes (in units 
of 10$^{-15}$~erg~cm$^{-2}$~s$^{-1}$) for the optical coronal and 
[\ion{O}{1}]~6300~\AA\ lines measured in the nuclear and adjacent  regions in the galaxy
  sample. \label{fwhmopt}}
\tablewidth{0pt}
\tablehead{
\colhead{}   & \colhead{} & \colhead{[Fe\,{\sc{vii}}]} & \colhead{}& \colhead{} & \colhead{} & \colhead{[O\,{\sc i}]} & \colhead{} & \colhead{} & \colhead{} & \colhead{[Fe\,{\sc x}]}
& \colhead{} & \colhead{} & \colhead{} & \colhead{[Fe\,{\sc xi}]} & \colhead{} \\
\cline{2-4}  \cline{6-8}  \cline{10-12}  \cline{14-16}
\colhead{Aperture} & \colhead{FWHM}  &  \colhead{$\Delta$V} & \colhead{Flux} & \colhead{} &  \colhead{FWHM}  &  \colhead{$\Delta$V}  & \colhead{Flux} & \colhead{}& \colhead{FWHM}  &
\colhead{$\Delta$V} & \colhead{Flux} & \colhead{} & \colhead{FWHM} & \colhead{$\Delta$V} & \colhead{Flux} \\
\colhead{(1)} & \colhead{(2)} & \colhead{(3)} &  \colhead{(4)} & \colhead{} &  \colhead{(5)} & \colhead{(6)} & \colhead{(7)} & \colhead{} &  \colhead{(8)} & \colhead{(9)} &
\colhead{(10)} & \colhead{} & \colhead{(11)} & \colhead{(12)} & \colhead{(13)}}
\startdata
\tableline
\multicolumn{16}{c}{Circinus} \\
\tableline
NUC     & 120  & -15  &  1.14$\pm$0.10 && 105 & 0.0  & 6.23$\pm$0.10 && 140 &  32 &  2.30$\pm$0.12 && 105 & -11  & 3.82$\pm$0.14 \\
        & 400  & -250 &  1.11$\pm$0.45 && ... & ...  &      ...      && ... & ... &   ...          && 290 & -130 & 1.83$\pm$0.36 \\
27~pc~N & 290  & -84  &  0.77$\pm$0.15 && 110 & -38  & 1.86$\pm$0.11 && 135 &  20 &  0.55$\pm$0.12 && 160 & -70  & 0.50$\pm$0.14 \\
53~pc~N & 130  & -79  &  0.20$\pm$0.06 && 135 & -76  & 0.85$\pm$0.12 && ... & ... &   ...          && ... & ...  &	...\\ 
27~pc~S & 160  &  20  &  0.49$\pm$0.10 && 105 &  24  & 2.37$\pm$0.10 && 120 &  60 &  1.07$\pm$0.10 && 105 & 0.0  & 1.36$\pm$0.11\\
        &  ... & ...  &   ...          && ... &  ... &    ...        && ... & ... &    ...         && 180 & -150 & 0.50$\pm$0.08 \\
\tableline
\multicolumn{16}{c}{NGC~1386} \\
\tableline
     
NUC     & 740  &  81  & 6.88$\pm$1.08 && 105 & -35  & 1.20$\pm$0.19 && 580 & -100 & 2.41$\pm$0.80 && 250  & -120 & 1.57$\pm$0.50 \\
78~pc~N & 220  &  290 & 0.70$\pm$0.21 && 575 &  87  & 1.87$\pm$0.50 && ... & ...  &   ...         && 320  & -45  & 0.69$\pm$0.11 \\
        & 1410 & -180 & 3.08$\pm$1.20 && ... & ...  &     ...       && ... & ...  &   ...         && ...  &  ... &    ... \\
78~pc~S & 700  &  173 & 1.03$\pm$0.08 && 105 & -43  & 3.40$\pm$0.16 && ... & ...  &   ...         && 280  & -105 & 0.59$\pm$0.34 \\
\tableline
\multicolumn{16}{c}{NGC 1068} \\
\tableline      
NUC      & 1670 & -39 & 142.0$\pm$17.3&& 900 &  0.0 & 173.9$\pm$9. && 1595 & -150 & 47.8$\pm$17.0 && 760 & -165 & 16.4$\pm$1.3 \\
105~pc~N & 1450 & -250& 81.0$\pm$11.5 && 1020& -50  & 177.1$\pm$8.4&& 1380 & -175 & 28.5$\pm$11.3 && 680 & -165 & 4.1$\pm$1.4 \\
210~pc~N & 990  & -695& 14.3$\pm$2.3  && 1280& -30  & 44.2$\pm$3.2 && ...  & ...  &  ...          && ... &  ... & ...         \\
105~pc~S & 1890 &  85 & 29.4$\pm$6.5  && 650 & -55  & 31.1$\pm$6.6 && 1800 &  220 & 11.7$\pm$6.5  && 705 & -115 & 8.0$\pm$0.8 \\
210~pc~S & 370  &  595&  1.6$\pm$0.2  && 720 & -75  &  7.8$\pm$0.5 && ...  &  ... &  ...          && ... & ...  & ...   \cr
\tableline
\multicolumn{16}{c}{NGC 3227} \\
\tableline 
NUC   &   965  & -125 & 4.09$\pm$0.56 && 520 & -35 & 20.04$\pm$0.9 && ...  &  ... &  ...          && ... & ...  & ...   \cr
\tableline
\multicolumn{16}{c}{MCG-6-30-15} \\
\tableline
NUC   &   160  &  -30 & 0.97$\pm$0.13 && 105 & -12 & 1.66$\pm$0.15 && 340  & -76  & 2.07$\pm$0.15 && 330 & -140 & 2.78$\pm$0.34 \\
      &  ç 670  &  -277& 1.21$\pm$0.30 && ... & ... &    ...        && 2090 & -340 & 4.33$\pm$1.00 &&1530 & -330 & 4.70$\pm$1.40 \\
\tableline
\multicolumn{16}{c}{NGC 3783} \\
\tableline
NUC      & 540  & -120 & 22.15$\pm$0.63&& 255 & -37& 16.87$\pm$0.43&& 625  & -170 &10.60$\pm$1.00 && 310 & -75  & 2.88$\pm$0.70 \\
         & 1380 & -600 & 7.80$\pm$1.57 && ... & ...&     ...       && 1320 & -635 & 7.85$\pm$2.02 && 1080& -320 & 7.69$\pm$2.37 \\
270~pc~N & ...  & ...  &    ...        && 310 & -39& 2.19$\pm$0.17 && 650  & ...  &    ...        && ... & ...  &  ...         \\
270~pc~S & ...  & ...  &    ...        && 295 & 0.0& 2.11$\pm$0.19 && 605  & ...  &    ...        && ... & ...  &  ...         \\
\enddata
\tablenotetext{a}{Two entries in a row gives the result of a two Gaussian fit to the 
line. Typical error in the Gaussian centroid is 20~\kms, but increases
largely (50-100\%) for low S/N lines}
\end{deluxetable}

\clearpage

\begin{table}
\begin{center}
\caption{FWHM and shifts  from the centroid position (both in km/s), and fluxes 
(in units of 10$^{-15}$~erg~cm$^{-2}$~s$^{-1}$) for the NIR coronal lines measured in the nuclear and adjacent regions of the galaxy
  sample\tablenotemark{a}. \label{fwhmnir}}
\begin{tabular}{lccccccc}
\tableline \tableline
& \multicolumn{3}{c}{[Si\,{\sc vi}]} & & \multicolumn{3}{c}{[Si\,{\sc vii}]} \\
\cline{2-4} \cline{6-8}
Aperture &  FWHM  &  $\Delta$V & Flux  & & FWHM  &   $\Delta$V & Flux \\
(1)      &   (2)  &       (3)        & (4)   & &  (5)  &       (6)         &  (7)\\
\tableline
\multicolumn{8}{c}{Circinus} \\
\tableline
NUC       &  125   & 	-37	 & 37.6$\pm$1.7 &&   100  &    18  & 53.0$\pm$0.4 \\  
          &  290   & 	 -210	 & 21.5$\pm$3.5 &&   270  &   -100 & 49.3$\pm$1.1 \\ 
17~pc~N   &  145   & 	 -60	 &  6.8$\pm$2.0 &&   100  &    20  & 1.9$\pm$0.2 \\ 
          &  245   & 	 -275	 & 0.4$\pm$0.2  &&   180  &    -90 & 1.7$\pm$0.2 \\ 
17~pc~S   &  140   & 	 -45	 & 12.2$\pm$2.7 &&   100  &    28  & 2.8$\pm$0.2 \\ 
          &  ...   & 	  ...	 & ...          &&   170  &   -85  & 4.40.6 \\ 
\tableline
\multicolumn{8}{c}{NGC~1068} \\
\tableline
NUC       &  950   & 	 -185	 & 421$\pm$90   &&   330  &    0.0 & 117.3$\pm$13.8 \\ 
    	  &  ...   & 	  ...	 & ...          &&   730  &   -250 & 131.0$\pm$30.0 \\ 
45~pc~N   &  175   & 	 -130	 & 48.2$\pm$5.7 &&   350  &    38  & 123.9$\pm$3.5 \\ 
          &  710   & 	 -180	 & 259.5$\pm$13 &&   710  &   -270 & 89.8$\pm$6.9 \\  
90~pc~N   &   650  & 	  -105   & 57.1$\pm$2.6 &&   390  &     35 & 31.7$\pm$1.4 \\ 
          &  ...   & 	 ...  	 & ...          &&   960  &  -580  & 27.3$\pm$3.4 \\ 
135~pc~N  &  790   & 	  47	 & 9.4$\pm$3.1  &&   ...  &  ...   & 26.0$\pm$2.4 \\ 
45~pc~S   &  1090  & 	 -38	 & 164$\pm$36   &&   290  &   -60  & 34.3$\pm$3.0 \\ 
90~pc~S   &  380   & 	  375	 & 25.5$\pm$3.0 &&   ...  &    ... & ... \\ 
135~pc~S  &  330   & 	  412	 & 11.6$\pm$1.3 &&   360  &    610 & 14.8$\pm$0.60\\ 
\tableline 
\multicolumn{8}{c}{NGC~3227} \\
\tableline       
NUC       &  660   & 	 -40	 & 25.0$\pm$4.0    &&   ...  &    ... & ... \\ 
\tableline
\end{tabular}
\tablenotetext{a}{A double entry in a row corresponds to the  results 
from a two Gaussian fit to the line.}
\end{center}
\end{table}

\clearpage

\clearpage

\clearpage

\clearpage

\clearpage

\clearpage

\clearpage

\clearpage

\clearpage

\clearpage

\clearpage

\clearpage

\clearpage

\clearpage

\clearpage

\end{document}